\documentclass[twocolumn]{aastex63}
\usepackage{float}

\def \refnu     {321.6\,MHz}
\def \refn      {C273$\alpha$}

\def \deg       {\text{$^{\circ}$}}
\def \amin      {\text{$^\prime$}}
\def \asec      {\text{$^{\prime\prime}$}}

\def \mjyb      {mJy~beam$^{-1}$}
\def \Hii       {\text{H\,\textsc{ii}}}

\def \Cii       {\text{[C\,\textsc{ii}]}}
\def \Hi        {\text{H\,\textsc{i}}}
\def \Htwo      {\text{H$_2$}}
\def \CO        {$^{12}$CO}
\def \treceCO   {$^{13}$CO}
\def \kms       {km~s$^{-1}$}
\def \cmc       {cm$^{-3}$}

\def \Msun      {\text{$\rm M_{\odot}$}}

\def \Av        {\text{A$_{\mathrm{V}}$}}

\received{...}
\revised{...}
\accepted{...}
\submitjournal{ApJ}
\shorttitle{Cool dark gas in Cygnus X}
\shortauthors{Emig et al.}

\begin{document}

\title{Cool dark gas in Cygnus X: The first large-scale mapping of low-frequency carbon recombination lines}

\correspondingauthor{Kimberly L. Emig}
\email{kemig@nrao.edu}

\author[0000-0001-6527-6954]{Kimberly L.~Emig}
\affiliation{National Radio Astronomy Observatory, 520 Edgemont Road, Charlottesville, VA 22903, USA}

\author{Pedro Salas}
\affiliation{Green Bank Observatory, 155 Observatory Road, Green Bank, WV 24915, USA}

\author{Loren D.~Anderson}
\affiliation{Department of Physics and Astronomy, West Virginia University,
Morgantown, WV 26506, USA}
\affiliation{Adjunct Astronomer at the Green Bank Observatory, P.O.~Box 2, Green
Bank, WV 24944, USA}
\affiliation{Center for Gravitational Waves and Cosmology, West Virginia
University, Chestnut Ridge Research Building, Morgantown, WV 26505,
USA}

\author{D.~Anish Roshi}
\affiliation{Florida Space Institute, University of Central Florida, Orlando 32826}

\author[0000-0002-0915-4853]{Lars Bonne}
\affiliation{SOFIA Science Center, USRA, NASA Ames Research Center, Moffett Field, CA 94 045, USA}

\author{Alberto D.~Bolatto}
\affiliation{Department of Astronomy, University of Maryland, College Park, MD 20742, USA}

\author{Isabelle A.~Grenier}
\affiliation{Universit\'e de Paris and Universit\'e Paris Saclay, CEA, CNRS, AIM, CEA Saclay, F-91190 Gif-sur-Yvette, France}

\author[0000-0003-2508-2586]{Rebecca C.~Levy}
\altaffiliation{NSF Astronomy and Astrophysics Postdoctoral Fellow}
\affiliation{Steward Observatory, University of Arizona, Tucson, AZ 85721, USA}
\affiliation{Space Telescope Science Institute, 3700 San Martin Drive, Baltimore, MD 21218, USA}

\author[0000-0002-4727-7619]{Dylan J.~Linville}
\affiliation{Department of Physics and Astronomy, West Virginia University, Morgantown, WV 26506, USA}
\affiliation{Center for Gravitational Waves and Cosmology, West Virginia University, Chestnut Ridge Research Building, Morgantown, WV 26505, USA}

\author[0000-0001-8061-216X]{Matteo~Luisi}
\affiliation{Department of Physics, Westminster College, New Wilmington, PA 16172, USA}
\affiliation{Center for Gravitational Waves and Cosmology, West Virginia University, Chestnut Ridge Research Building, Morgantown, WV 26505, USA}

\author[0000-0002-2862-307X]{M.~Riley Owens}
\affiliation{Department of Physics, University of Cincinnati, Cincinnati, OH 45221, USA}

\author{J.~Poojapriyatharsheni}
\affiliation{Department of Physics, Lady Doak College, Madurai, Tamil Nadu, 625002, India}

\author[0000-0003-3485-6678]{Nicola Schneider}
\affiliation{I. Physikalisches Institut, Universität zu Köln, Zülpicher Str. 77, 50937 Köln, Germany}

\author[0000-0001-7523-570X]{Luigi Tibaldo}
\affiliation{IRAP,Universit\'e de Toulouse, CNRS, CNES, 9 avenue Colonel Roche, 31028 Toulouse, Cedex4, France}

\author{Alexander G.~G.~M.~Tielens}
\affiliation{Department of Astronomy, University of Maryland, College Park, MD 20742, USA}
\affiliation{Leiden University, P.O.Box 9513, NL-2300 RA, Leiden, The Netherlands}

\author{Stefanie K.~Walch}
\affiliation{I. Physikalisches Institut, Universität zu Köln, Zülpicher Str. 77, 50937 Köln, Germany}

\author{Glenn J.~White}
\affiliation{School of Physical Sciences, The Open University, Walton Hall, Milton Keynes, MK7 6AA, UK}
\affiliation{RAL Space, STFC Rutherford Appleton Laboratory, Chilton, Didcot, Oxfordshire, OX11 0QX, UK}

%%%%%%%%%%%%%%%%%%%%%%%%%%%%%%%%%%%%%%%%

\begin{abstract}

Understanding the transition from atomic gas to molecular gas is critical to explain the formation and evolution of molecular clouds. However, the gas phases involved, cold \Hi{} and CO-dark molecular gas, are challenging to directly observe and physically characterize. 
We observed the Cygnus X star-forming complex in carbon radio recombination lines (CRRLs) at 274--399 MHz with the Green Bank Telescope at 48\amin{} (21 pc) resolution. 
Of the 30 deg$^2$ surveyed, we detect line-synthesized \refn{} emission from 24 deg$^2$ and produce the first large-area maps of low-frequency CRRLs, which likely originate in CO-dark molecular gas. 
The morphology of the \refn{} emission reveals arcs, ridges, and extended possibly-sheet-like gas which are often found on the outskirts of CO emission. We find a correlation between velocity-integrated \refn{} and the 8~$\mu$m intensity with a power-law slope of $1.3 \pm 0.2$. 
We interpret the relation as the dependence of cool dark gas emission on the FUV radiation field, $G_0 \approx 40 - 160$.
We determine the typical angular separation between \refn{} and \treceCO{} emission to be 12 pc. 
Velocity differences between \refn{} and \treceCO{} are apparent throughout the region and have a typical value of 2.9~\kms{}.  
We estimate gas densities of $n \approx 20 - 900$~\cmc{} with a nominal $n \approx 400$~\cmc{} in the C$^{+}$/H$_2$ layer.
The evolution of the \refn{} gas seems to be dominated by turbulent pressure, with a characteristic timescale to form \Htwo{} of about 2.6 Myr. These observations underline the richness of low-frequency CRRLs to provide revelatory insights into the characteristics of (CO-)dark gas and the evolution of molecular gas.

\end{abstract}


%%%%%%%%%%%%%%%%%%%%%%%%%%%%%%%%%%%%%%%%

\section{Introduction} 
\label{sec:intro}

Understanding the transition from atomic gas to molecular gas is critical to explain the formation and evolution of molecular clouds. There are two main phases of gas directly involved in the \Hi{}-to-\Htwo{} transition, cold \Hi{} and CO-dark molecular gas. They are often referred to as cool ``dark gas'', since they are challenging to directly observe with typical tracers, \Hi{} 21 cm and CO rotational transitions, leading to a dearth of knowledge around the formation of molecular gas in galaxies. Cool dark gas is estimated to make up a considerable fraction of the Galaxy's interstellar medium (ISM) mass \citep{Grenier2005, Remy2018, Busch2021, Murray2020, Marchal2024}. 

On cloud-scales, the \Hi-to-\Htwo{} transition marks where gas is converted from a mostly atomic state to mostly molecular. \Htwo{} formation occurs via catalytic reactions on the surfaces of interstellar dust grains \citep[for a review, see][]{Wakelam2017}. The gas density, far-ultraviolet (FUV; 6-13.6~eV) radiation, and dust properties describe the column densities (or $A_V$) where and how rapidly this transition occurs,  typically between $A_V =0.4-3$ from models \citep[e.g.,][]{vanDishoeck1986, Wolfire2010, Sternberg2014, Imara2016, Schneider2023}. 
The ambient radiation field in the ISM rapidly photo-dissociates \Htwo{} and also heats the gas through photoelectric heating. Dust helps to shield and attenuate FUV radiation.
When the \Htwo{} opacity to FUV radiation becomes high enough such that self-shielding of \Htwo{} is efficient, the abundance of \Htwo{} rapidly increases. Because carbon has a lower ionization potential (IP = 11.26~eV) than hydrogen (IP = 13.6~eV), carbon may be singly ionized in regions where hydrogen is predominantly molecular. CO, the workhorse tracer of molecular clouds \citep[e.g.,][]{Bolatto2013b, Heyer2015}, reaches the abundances required to self-shield, and thus become observable, only deeper into the cloud.

% properties of cold, dark gas - HI view
The gas phases involved in the \Hi{}-to-\Htwo{} transition are difficult to observe. 
\Hi{} from warm ($T \sim 7000$~K) gas dominates the main \Hi{} 21 cm observable, and cold \Hi{} \citep[$T \sim 70$~K;][]{Heiles2003b} is observable towards select (i.e., often nearby, high-latitude) lines-of-sight, or through \Hi{} self-absorption (HISA) when illuminated by a warm-\Hi{} background component \citep{Heeschen1955}. These studies have provided foundational insights into diffuse and translucent cloud conditions \citep{McClure-Griffiths2023}. To infer the presence of cool dark gas and investigate its properties, decomposing far-IR dust emission \citep[e.g.,][]{Planck2011o}, gamma-ray emission \citep[e.g.,][]{Grenier2005}, and \Cii{} 158~$\mu$m emission \citep[e.g.,][]{Pineda2013, Tang2016} into the multi-phase ISM components from which they arise has also been employed. Recently strides have been made in directly observing the dark components, with \Cii{} especially in star forming environments \citep{Beuther2014,Schneider2023,Bonne2023} and with quasi-thermal OH emission \citep{Busch2019, Busch2021}. Carbon radio recombination lines (CRRLs) at \textit{low-frequencies} ($< 1 $~GHz) strongly complement these dark gas probes.  

Generally speaking, CRRLs arise from high principal quantum number ($\mathsf{n}$) transitions in C$^+$ gas, where carbon is predominantly singly ionized. At low radio frequencies ($<$1~GHz, $\mathsf{n} \gtrsim 187$), CRRL emission is enhanced due to stimulation\footnote{With stimulated emission, the level populations of atoms do not reflect the Boltzmann distribution due to interactions of the electrons with the radio continuum.} and dielectronic capture\footnote{\cite{Watson1980} and \cite{Walmsley1982} showed that at low temperatures ($T_e \lesssim 100$~K) electrons can recombine with carbon ions at high $\mathsf{n}$ states by simultaneously exciting the \Cii{} $^2 P_{1/2} - ^2 P_{3/2}$ fine structure line at 158~$\mu$m, a process known as dielectronic capture.} \citep{Shaver1975a, Watson1980}. Atomic physics modeling shows that low-frequency CRRLs trace gas with temperatures of 20--100~K and densities $n_e \approx 0.01 - 0.1$~\cmc{} ($n_{\mathrm{H}} \approx 100 - 1000$~\cmc{}) \citep{Walmsley1982, Payne1994, Salgado2017a}. 
The warmer and denser conditions of C$^+$ gas found in classic, dense photo-dissociation regions (PDRs) are observed with high-frequency CRRL emission but not at \textit{low-frequencies} due to increased pressure broadening, decreased stimulation and fainter background continuum, and optically thick free-free continuum of associated regions.

% state of the field - historical / low-res view
Low-frequency CRRL observations are excellent probes of the \Hi{}-to-\Htwo{} transition because they trace the density and temperature regime where the transition takes place. However, they have largely been underutilized due to a lack of high-resolution and high-sensitivity telescopes at the relevant frequencies. Pioneering work with low-frequency CRRLs have shown that they are ubiquitous in large (2\deg{}--110\deg{}) beams where background continuum is bright \citep{Anantharamaiah1985a, Erickson1995, Roshi1997, Kantharia2001, Roshi2002, Vydula2023}, for example towards the Inner Galaxy. \citet{Roshi2002} found CRRL emission at 327 MHz to resemble the radial extent of intense \CO{} emission in our Galaxy. \citet{Roshi2011} used Ooty Telescope 327 MHz survey data associated with the Riegel-Crutcher Cloud, identified that the narrow CRRL components are coincident with HISA features, estimated \Htwo{} formation rates which far exceeded dissociation rates, and thereby showed that CRRLs trace gas in the process of forming molecular gas.

\begin{figure*}[!t]
    \centering
    \includegraphics[width=0.49\textwidth]{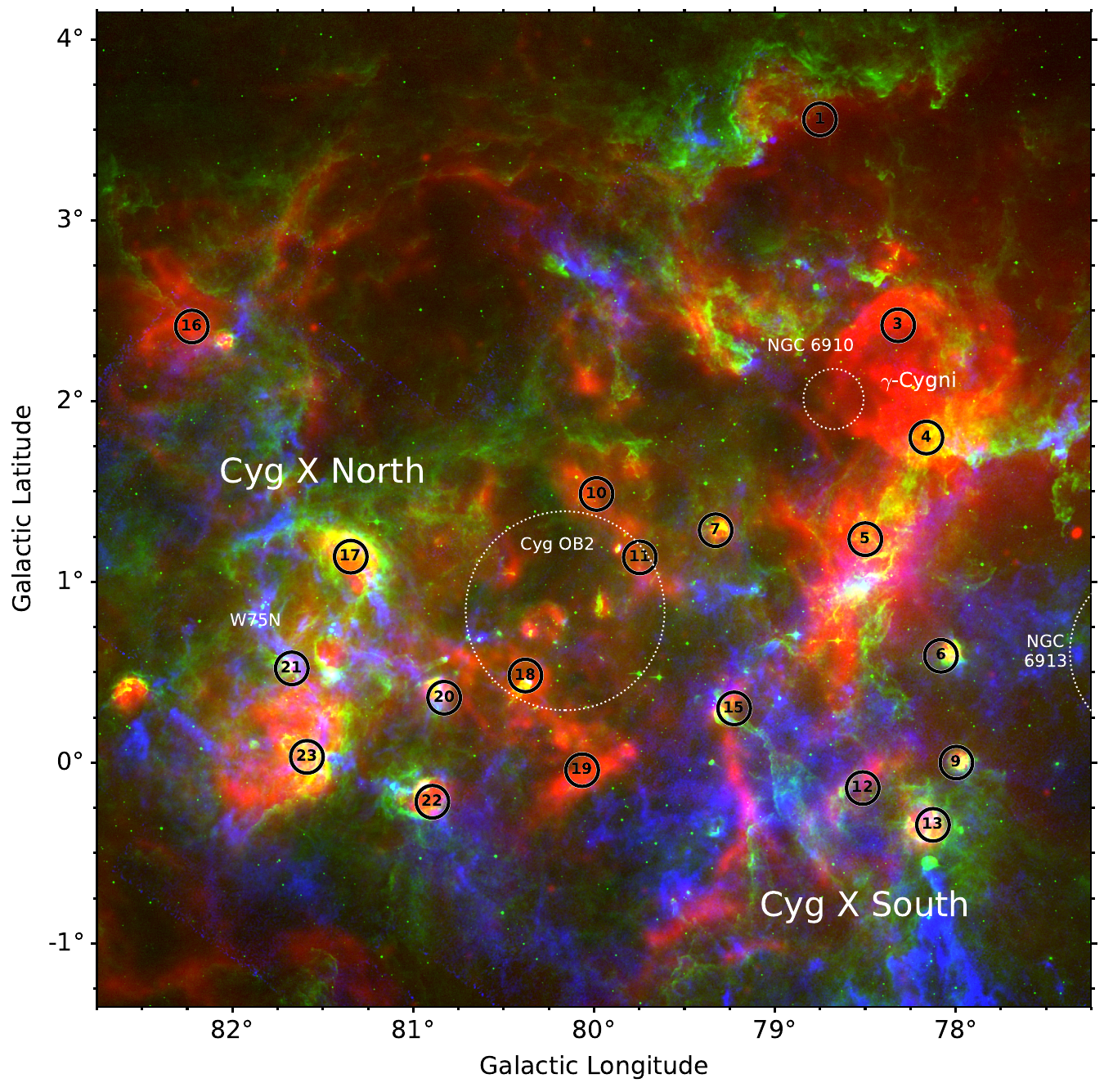}
    \includegraphics[width=0.49\textwidth]{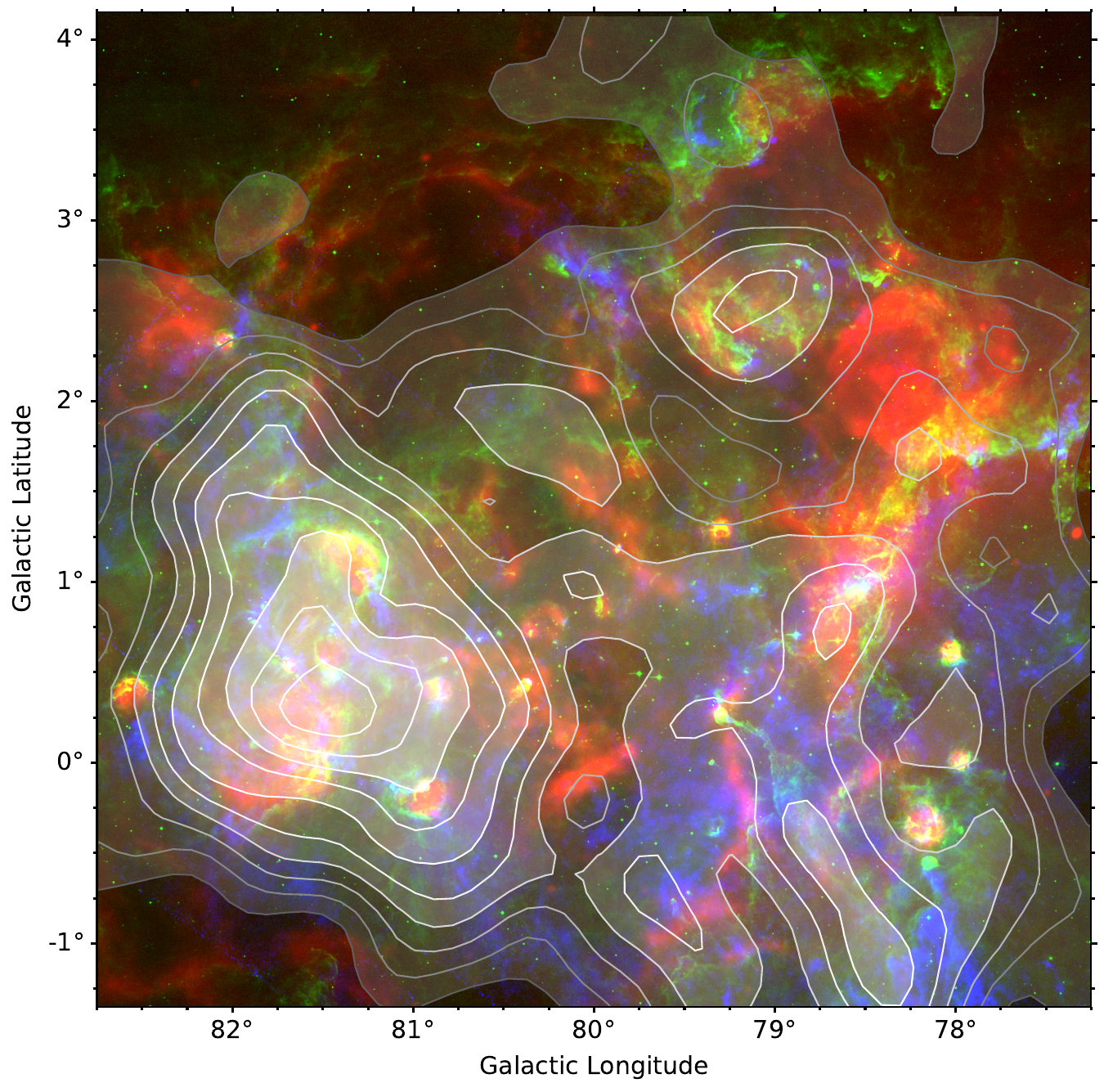}
    \caption{RGB compilation in the Cygnus X region covering the footprint that we surveyed; in red is the CGPS 1.4\,GHz continuum \citep{Taylor2003}, green is MSX 8~$\mu$m PAH emission \cite{Schneider2006}, and blue is \treceCO{} (1-0) tracing (some of the) molecular gas \citep{Schneider2010}.
    \textit{Left:} The numbers enclosed by a circle in black indicate the `DR' continuum sources identified in 5 GHz observations by \cite{Downes1966}, which are mostly thermal \Hii{} regions, except for DR3 and DR4 that make up the supernova remnant $\gamma$-Cygni. Dotted circles show the Cyg OB2 association and open clusters NGC 6910 and NGC 6913. The region at $\ell > 80$\deg{} commonly referred to as Cyg X North and at $\ell < 80$\deg{} (also typically at lower latitudes) as Cyg X South are also indicated. 
    \textit{Right:} Contours of velocity-integrated \refn{} (moment 0), drawn at $[3,5.5,8,...28]\sigma$. \refn{} is detected from 75\% of the mapped region.}
    \label{fig:cyg_intro}
\end{figure*}

% state of the field - recent, higher res
Recently, higher resolution (arcsec to arcmin) studies of low-frequency CRRLs have been enabled, thanks to upgraded receivers and high-resolution telescopes at low frequencies \citep{Salas2017, Oonk2017, Salas2018, Salas2019, Chowdhury2019, Roshi2022}. Detailed studies of gas in the Perseus Arm along the line-of-sight towards the Cassiopeia A (Cas A) supernova remnant find CRRL emitting layers to trace the surface of a molecular cloud \citep{Oonk2017, Salas2018}. \cite{Chowdhury2019} used GMRT 430 MHz observations and find clumps of CRRL emission on scales of $< 0.3$~pc embedded in larger-scale ($\geq 7$~pc) diffuse emission. In the Orion star-forming region, CRRLs used in conjunction with \Cii{}~158~$\mu$m provided key physical properties to anchor models of photodissociation regions \citep{Salas2019}.

% state-of the field -- prior mapping
Although large-beam surveys are highly valuable, they have so far not produced \textit{maps} of low-frequency CRRL emission. Only gas in front of the extremely bright, $S_{100~\mathrm{MHz}} \sim 10^4$~Jy, Cas A has been resolved and mapped over the 8\amin{} diameter (0.014 deg$^2$ area) supernova remnant \citep{Kantharia1998a, Salas2018, Chowdhury2019}. 

In this article, we present the results of CRRL observations at 274--399 MHz using the Green Bank Telescope (GBT) in a 30~deg$^2$ area covering the Cygnus X star forming region. We use these observations to investigate the \Hi{}-to-\Htwo{} transition (or vice versa) in the formation and/or destruction of molecular clouds. To our knowledge, this is the first mapping of CRRLs arising from cold, diffuse gas that is larger than $0.014$~deg$^2$. The high surface brightness of the radio continuum elevates the line intensities of the stimulated CRRLs. The gas content and stellar activity in Cygnus~X allows us to characterize the \Hi{}-to-\Htwo{} transition in actively forming molecular gas \citep{Schneider2023, Bonne2023} and in the presence of an elevated radiation field.

%%%%%%%%%%%%%%%%%%%%%%%%%%%%%%%%%%%%%%%%%

\section{Overview of the region}
\label{sec:cygx}

Cygnus~X is a nearby (approximately 1.5~kpc) massive star-forming complex that spans more than 6 degrees in size (see Figure~\ref{fig:cyg_intro}). Cygnus~X hosts more than 170 massive OB stars \citep{LeDuigou2002, Comeron2012, Wright2015, Berlanas2018, Comeron2020, Quintana2021} --- some of which are surrounded by bright \Hii{} regions \citep[e.g.,][see `DR' source IDs in Figure~\ref{fig:cyg_intro}]{Downes1966} --- a large number of actively forming stars \citep{Motte2007, Beerer2010, Bontemps2010, Ortiz-Leon2021}, and stellar remnants of supernovae and pulsars \citep[e.g.,][]{Ladouceur2008}. Cyg OB2 \citep[$d \sim 1.4-1.7$~kpc;][]{Berlanas2019, Cantat-Gaudin2020, Quintana2021}, a prominent association of stars in the heart of the region, has a stellar mass of around $2 \times 10^4$~\Msun{} \citep{Wright2015} and age 3--5~Myr \citep{Wright2010, Berlanas2020}. Cyg OB2 bathes the region in a high UV radiation field \citep[$G_0 \gtrsim 5-1000$; e.g.,][]{Schneider2016b} \footnote{$G_0$ indicates the FUV-field (6--13.6 eV) expressed in units of a one-dimensional \citet{Habing1968} interstellar field of $1.6 \times 10^{-3}$ erg cm$^{-2}$ s$^{-1}$.} and has had profound impacts by triggering star formation \citep{Schneider2016b, Deb2018}, photo-evaporating cold clouds \citep{Wright2012, Emig2022}, and through its stellar winds \citep{Ackermann2011a, Abeysekara2021}.

The Cygnus~X region contains an abundance of molecular gas \citep[e.g.,][]{Schneider2006}, with two main concentrations of emission generally referred to as Cyg X North ($M_{\mathrm{H_2}} \approx 3 \times 10^5$~\Msun) and Cyg X South ($M_{\mathrm{H_2}} \approx 5 \times 10^5$~\Msun), with a cleared medium in between aligned with Cyg OB2 (see Figure~\ref{fig:cyg_intro}). The molecular clouds in Cyg X North that are primarily associated with DR21 and W75N may be interacting \citep{Dickel1978, Dobashi2019, Schneider2023, Bonne2023}; a cloud-cloud collision has also been hypothesized for clouds in Cyg X South \citep{Schneider2006}. A foreground cloud, part of the Great Cygnus Rift \citep[$d \sim 600-800$~pc; e.g., see review in][]{Uyaniker2001}, also contributes to some emission in this direction.

In this article, we assume the distance to the Cygnus X clouds is $1.5 \pm 0.2$~kpc \citep[e.g.,][]{Rygl2012}, for which 1\amin{}~$= 0.44 \pm 0.06$~pc.

\section{Data}    
\label{sec:data}

%%%%%%%%%%%%%%%

\subsection{GBT Observations and Data Reduction}

% Pedro
We mapped a $5.5 \times 5.5$ sq.~degree ($144 \times 144$ sq.~pc) region centered on $(\ell,b)=(80\deg,1.4\deg)$ using the $342$~MHz prime focus receiver (Rcvr\_342) on the 100~m Robert C. Byrd Green Bank Telescope \citep[GBT;][]{Prestage2009}. The observations were carried out between April 16, 2021 and May 23, 2021 as part of project GBT21A-292, sessions 14 to 23. Radio recombination line transitions C255$\alpha$ through C282$\alpha$ at 292 -- 394~MHz were covered.

The observations used the Versatile GBT Astronomical Spectrometer \citep[VEGAS;][]{Prestage2015} in spectral line mode to transform the raw voltages into spectra. We observed using the total power mode, firing a noise diode of $\approx10\%$ of the receiver temperature ($20$--$70$~K) every other integration. We split the frequency range covered by the receiver into seven spectral windows, each $23.44$~MHz wide and with $2^{15}$ channels $0.7$ kHz wide (VEGAS mode $10$). We recorded the linear orthogonal auto-cross correlation products, XX \& YY, and used an integration time of $\approx1.9$~s.

At the start of observing sessions 14, 15, 16 and 19, between April 16, 2021 and May 21, 2021, we determined pointing corrections by observing a bright point-like 3C source (3C295 or 3C48). In general the pointing corrections were smaller than $1\arcmin$, less than $3\%$ of the half power beam width at the highest RRL frequency observed ($31\arcmin$ at $399.14$~MHz for the C$254\alpha$ RRL). During these same sessions we also observed the bright point-like 3C source using position switching to determine the equivalent temperature of the noise diode. We use the flux density scale of \citet{Perley2017} and adopt an aperture efficiency of $0.71$ for the GBT. Given the small pointing offsets and stability of the temperature of the noise diodes, we decided not to derive pointing corrections nor observe a flux calibrator during other observing sessions. 

To calibrate the data we used custom \texttt{Python} data reduction scripts. We follow the formalism described in \citet{Winkel2012}, that is, we perform a frequency dependent calibration, as opposed to the default behavior offered by \texttt{GBTIDL} \citep{Marganian2013}. 

The first step in our data reduction is to find the gain, including a second order term \citep[see e.g.,][]{Salas2019}, using continuum maps for the region. The continuum maps are derived for the central frequency of each spectral window using the methods described in \citet{Emig2022}. Then, we split each spectral window into $1000$~km~s$^{-1}$ sub-windows centered on the hydrogen radio recombination lines (HRRLs). We calibrate each sub-window to antenna temperature applying the previously derived gain, and removing the contribution from the noise diode for the integrations where it was on. Then, we remove the continuum and baseline by fitting an order $11$ polynomial to line free-channels. The line free-channels are defined as being more than $50$~km~s$^{-1}$ away from the brightest RRL in each sub-window, the HRRLs. An order 11 polynomial captures most of the baseline structure, although in some cases a lower order would have sufficed. To remove radio frequency interference (RFI) we run \texttt{AOFlagger} \citep{Offringa2012} on each continuum subtracted sub-window. This calibration is performed for each observing session and by treating each polarization independently.

The next step in our data reduction is line stacking. We start by selecting the sub-windows, i.e., lines, that will make it into a stack. For each observing session, we visually inspect the calibrated spectra, one for each CRRL and polarization, and select those that show a smooth bandpass (i.e., can be modeled using a polynomial), show no significant leftover RFI, and have less than $30$\% of the data flagged. The selected lines are interpolated to a common velocity grid, with $1200$ channels $0.5$~km~s$^{-1}$ wide. The interpolated CRRL spectra for a single polarization are averaged together using $T_{\rm{sys}}^{2}/\Delta t$ as weights, with $T_{\rm{sys}}$ the system temperature and $\Delta t$ the integration time. We compare the stacks in both polarizations and look for any spurious features. If the stacks in both polarizations agree, then we repeat the stacking process using both polarizations. We found no instances where both polarizations disagreed by more than their noise. After this step we are left with one set of averaged CRRL spectra for each observing session.

We gridded the averaged CRRL spectra for each observing session using the \texttt{gbtgridder}\footnote{we use version 2.0 of the \texttt{gbtgridder} (\url{https://github.com/GreenBankObservatory/gbtgridder/tree/release_2.0}) which is a wrapper for GBT data around \texttt{cygrid} \citep{Winkel2016b}.}. During the gridding process we use a Gaussian function as the interpolation kernel with a width equal to the half power beam width of the GBT at the frequency of the lowest CRRL included in the stacks. Finally we averaged together all the cubes from the different observing sessions. This results in a single CRRL cube, which also contains HRRL emission.

We then divide the line intensity, $T_{\mathrm{L}}$, at each voxel of the cube by the intensity of the continuum, $T_{\mathrm{C}}$, creating a line-to-continuum ratio, $T_{\mathrm{L}} / T_{\mathrm{C}}$, data cube. 
We constructed the continuum image at \refnu{} following the methods described in \cite{Emig2022}. As we describe in Section~\ref{sec:crrl_emission}, CRRLs dominated by stimulated emission have an optical depth equal to the line-to-continuum ratio, $\tau_{\mathrm{L}} \approx - T_{\mathrm{L}} / T_{\mathrm{C}}$, resulting in the line-to-continuum ratio being directly proportional to the physical quantities of interest \citep{Shaver1975a, Salgado2017b}. We use the $T_{\mathrm{L}} / T_{\mathrm{C}}$ data cube to present our observational results.

The line-synthesized data cube has an effective frequency of \refnu{} corresponding to an effective principal quantum number of \refn{}. The beam FWHM is 48.3\amin{} and the typical noise is $\sigma_{T_L / T_C} = 1.9 \times 10^{-4}$ with a 0.5~\kms{} channel resolution. We constructed a 3D estimate of the noise at each voxel, described in Appendix~\ref{ap:sec:noise}.

Throughout this article, we analyze results from the $T_{\mathrm{L}} / T_{\mathrm{C}}$ data cube and often refer to this simply as \refn{} emission.

%%%%%%%%%%%%%%%

\subsection{Ancillary Data}

\textit{\treceCO}. We compare \refn{} emission with a bulk tracer of molecular gas, \treceCO{} (1--0) at 110.20~GHz. The opacity of \treceCO{} is less than \CO{} (1-0) and with the deep sensitivity of the observations ($\sim 0.25$~K), the emission is sensitive to even the low column densities of outer cloud layers. Data were kindly provided by the Milky Way Imaging Scroll Painting (MWISP) project \citep{Su2019, Zhang2024}. The Purple Mountain Observatory (PMO) observations cover the entire region mapped by our GBT observations at 15\amin{} angular and 0.17~\kms{} velocity resolutions. 

In Figure~\ref{fig:cyg_intro}, we also show high resolution (48\asec{}) \treceCO{} mapped by the Five College Radio Astronomy Observatory (FCRAO) 14~m telescope \citep{Schneider2010, Schneider2011} with a noise of 0.2~K at 0.1~\kms{} channel resolution.

\textit{\CO}. We compare \refn{} emission with \CO{} (1--0) emission at 115.27~GHz from molecular gas. We use \CO{} observations mapped over our full survey region by \cite{Leung1992} and \cite{Dame2001} with the Center for Astrophysics Millimeter-Wave Telescope. These \CO{} data have a native beam size of 8.7\amin{} and noise of 0.12~K at 0.65~\kms{} channel resolution.

\textit{8~$\mu$m}. 8~$\mu$m emission mainly traces UV-heated small grains and polycyclic aromatic hydrocarbons (PAHs) in PDRs where the gas is typically in an atomic state. In Figure~\ref{fig:cyg_intro}, we compare Midcourse Space Experiment \citep[MSX;][]{Price2001} 8.3~$\mu$m emission that has an angular resolution of 20\asec{} \citep[see][]{Schneider2006}.

\textit{\Hi{} 21 cm}. Spectra of \Hi{} 21~cm emission are obtained from the HI4PI Survey with the Effelsberg telescope at 16.2\amin{} resolution and 43~mK sensitivity in 1.3~\kms{} channels \citep{HI4PICollaboration2016}.

\textit{RRLs at 5.8~GHz}. 5.8~GHz RRL observations from the GBT taken with the same observational setup and data reduction as that of the GBT Diffuse Ionized Gas Survey \citep[GDIGS;][]{Anderson2021} are used to compare RRL intensities at different frequencies.  The data have a native spatial resolution of 2.65\amin{} and a spectral resolution of 0.5~\kms{}.  Compared to GDIGS, the Cygnus X data were taken with less time per pointing, resulting in spectral noise of 28~mK.

\textit{1.4 GHz continuum from CGPS.} We plot 1.420~GHz continuum emission in this region as observed by the Canadian Galactic Plane Survey \citep[CGPS][]{Taylor2003} in Figure~\ref{fig:cyg_intro}. We convolved and stitched the survey data products as in \citet{Emig2022} to a common resolution of 2\arcmin{}. The standard deviation in a relatively low emission region of the image is $\sigma = 0.03$~K (0.7~\mjyb{}).

%%%%%%%%%%%%%%%%%%%%%%%%%%%%%%%%%%%%%%%%%

\section{Description of Low-Frequency Carbon Recombination Line Emission}
\label{sec:crrl_emission}

The solution to the radiative transfer equation for the brightness of a CRRL \citep{Shaver1975a}, from upper level $n+1$ to lower level $n$, i.e., an $\alpha$ transition for which $\Delta n = 1$, is, in the optically thin limit:
\begin{equation}
    T_L \approx \tau^{*}_L \left( b_{n+1} T_e - b_{n}\beta_{n} T_C \right)
\label{eq:tlb}
\end{equation}
where {$T_{L}$ is the C$n\alpha$ line temperature, $T_e$ is the electron temperature of the emitting gas, $T_C$ is the continuum background temperature, $b_{n(+1)}$ and $\beta_n$\footnote{$\beta_n$ is the correction factor for stimulated emission as defined by \citet{Brocklehurst1972}, $\beta_n = \frac{1 - (b_{n+1}/b_n)\exp(-h\nu/k T_e)}{1 - \exp(-h\nu / k T_e)}$.} are the departure coefficients which measure the deviation of the level populations from LTE\footnote{LTE refers to the level populations being described by a Boltzmann distribution.} values, and $\tau^{*}_L$ is the LTE line optical depth as
\begin{equation}
\tau^{*}_L = 2.042 \times 10^{6} \left( \frac{EM_{C^+}}{\rm{cm^{-6}~pc}} \right) \left( \frac{T_e}{\rm{K}} \right)^{-2.5} \left( \frac{\rm{Hz}}{\Delta \nu } \right)
\label{eq:tau_lte}
\end{equation}
where $EM_{\mathrm{C^+}}$ is the emission measure $EM_{\mathrm{C^+}} = \int n_e n_{\mathrm{C^+}} d\ell$ and $\Delta \nu$ is the line width.

Spontaneous emission, the $b_{n+1}T_e$ term in Equation~\ref{eq:tlb}, typically dominates high-frequency CRRLs, where background continuum, $T_C$, is faint and the $b_n \beta_n$ coefficients are small. Stimulated emission, the $b_n \beta_n T_C$ term in Equation~\ref{eq:tlb}, typically dominates low-frequency CRRLs, where both $T_C$ and the departure coefficients, $b_n \beta_n$, take on large values.

When the background continuum term dominates and the CRRL emission is primarily stimulated, Equation~\ref{eq:tlb} becomes,
\begin{eqnarray}
T_L \approx - \tau^{*}_L b_n \beta_n T_C \nonumber 
\\
\approx - 
\tau_L T_C
\label{eq:tl_stim}
\end{eqnarray}
where $\tau_L$ is the observed non-LTE optical depth, and arriving at the standard relation \citep{Salgado2017b},
\begin{equation}
\frac{\int T_L \Delta \nu }{T_C} \approx   2.042 \times 10^{6} \, \mathrm{Hz} \, (-b_n \beta_n) \left( \frac{EM}{\rm{cm^{-6}~pc}} \right) \left( \frac{T_e}{\rm{K}} \right)^{-2.5} 
\label{eq:lcratio}
\end{equation}
where the departure coefficients, $b_n \beta_n$, take on negative values for lines observed in emission and are themselves dependent upon electron temperature, density, and the radio-continuum radiation field \citep[e.g.,][]{Salgado2017a}.

%%%%%%%%%%%%%%
\begin{figure}
    \centering
    \includegraphics[width=0.47\textwidth]{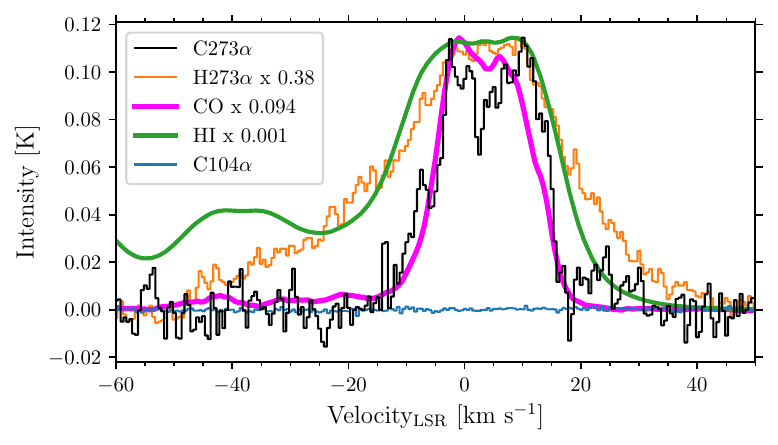}
    \caption{Spatially-averaged line emission of the survey region. The CRRL 322 MHz temperature brightness spectrum of the effective \refn{} line, which is the spatial average over the entire mapped region. Spatially-averaged spectra of additional gas phases have been normalized to match the peak of \refn{}; \CO{} \citep{Dame2001} traces the molecular phase, \Hi{} 21\,cm \citep{HI4PICollaboration2016} traces the atomic phase, and H273$\alpha$ (these data) traces diffuse ionized gas. We also show the non-detection of C$104\alpha$ from GDIGS 5.8 GHz observations \citep{Anderson2021}, emphasizing the stimulated nature of the \refn{} \refnu{} observations.}
    \label{fig:spec_spatialavg}
\end{figure}

Figure~\ref{fig:spec_spatialavg} shows the spatially-averaged line temperature of the 322~MHz \refn{} in the survey region. \refn{} peaks at about $T_L \approx 110$~mK. In comparison, the spectrum of 5.8~GHz CRRLs, effectively C104$\alpha$, extracted from the same area in GDIGS observations \citep{Anderson2021} is not detected with a 3$\sigma$ upper limit of $T_L < 1.8$~mK. Stimulated line emission is directly proportional to the continuum temperature (Equation~\ref{eq:tl_stim}), and in this region, the continuum is largely $T_C \propto \nu^{-2.1}$, \citep{Wendker1991, Xu2013, Emig2022}, except towards the supernova remnant where steeper indices are observed.
For stimulated emission, the expected line temperature at 5.8 GHz would be $\simeq$0.014~mK, as calculated by $T_L(5.8\,\mathrm{GHz}) \propto T_L(322~\mathrm{MHz}) \cdot \nu^{-3.1} \approx 110~\mathrm{mK} \cdot (5800/322)^{-3.1} \approx 0.014$~mK, which is consistent with the GDIGS non-detection of $T_L(5.8\,\mathrm{GHz}) < 1.8$~mK. Note the additional $\nu^{-1}$ dependence from the line optical depth for a Doppler-broadened line profile. Whereas for spontaneous emission, $T_L \propto T_e \nu^{-1}$ and the line temperature at 5.8 GHz is expected to stay within a factor of two of 6.1~mK, for $b_n$ values that are typically between 0.3--2 \citep{Salgado2017a}. This is inconsistent with the observed line intensities. The spectral line energy distribution (SLED) of the CRRLs therefore indicates that the 322 MHz CRRLs are dominated by stimulated emission.

%%%%%%%%%%%%%%%%%%%%%%%%%%%
\section{\refn{} Emission Properties}   
\label{sec:results}

The observed \refn{} emission is likely dominated by stimulated emission (Section~\ref{sec:crrl_emission}) and is therefore described by Equations~\ref{eq:tl_stim}~\&~\ref{eq:lcratio}. Since the line-to-continuum ratio is directly proportional to the physical properties of the emission, we present the results in terms of a line-to-continuum ratio $T_{\mathrm{L}} / T_{\mathrm{C}}$ data cube throughout the paper, as is commonly done for low-frequency CRRLs \citep[e.g.][]{Kantharia2001, Roshi2002}. While we may refer to the emission simply as \refn{} emission, it should be taken to mean  $T_{\mathrm{L}} / T_{\mathrm{C}}$.

%%%%%%%%%%%%%%%%%%%%%%%%
\subsection{Velocity-Integrated \refn{} Map}

In Figure~\ref{fig:cyg_intro}, we show the velocity-integrated \refn{} emission (Moment 0) that has been integrated over -10 to 14~\kms{}. The velocity-integrated \refn{} emission shows that \refn{} is detected throughout most of the region, having emission with a significance greater than $3\sigma$ ($5\sigma$) over 24.1 (20.3) deg$^2$. The brightest \refn{} emission is coincident with Cyg X North, the region of the highest star formation rate surface density that is young and active. Elongated structures as well as somewhat localized enhancements of emission are apparent. We show maps of overlays of \treceCO{}, 8~$\mu$m, and low-frequency continuum at matched spatial resolution in the Appendix Figure~\ref{ap:fig:overlay_matched_res}.

We do not show maps of the intensity-weighted central velocity (Moment 1) or the intensity-weighted velocity dispersion (Moment 2). The data have relatively low signal-to-noise ratios and thus do not produce reliable and meaningful higher order Moment calculations \citep{Teague2019}.

%%%%%%%%%%%%%%%%%%%%%%%%%
\subsection{\refn{} Channel Maps}

Channel maps of \refn{} are shown in Figure~\ref{fig:chan_CRRL}. For visual aid, we mark the locations of well-known radio continuum sources, first cataloged by \cite{Downes1966} at 5~GHz and 10.8\amin{} resolution. 
\refn{} emission appears both extended and elongated throughout most of the channel maps.  
The size scales of emission range from a fraction of a beam, $\sim$16 \amin{} (8 pc), to resolved extensions more than 3\deg{} (80 pc) across. 

Ridges and arcs can be seen in Channel 1.3~\kms{} surrounding DR 9/12/13, in 3.3~\kms{}, in 6.3~\kms{}, in 7.3~\kms{} bridging DR 22 and 23, in 8.3~\kms{} upwards from DR 20, and in 11.3~\kms{} forming an arc in the Western half of the map.  Notably, there is an elongated ridge of emission, peaking coincidentally with DR15 in the 0.3~\kms{} channel map.  At velocities $\gtrsim 3$~\kms{}, Cyg X North dominates the brightest emission in the region, most prominently overlapping spatially with DR 17, 20, 21, 22, and 23. Emission in this region also appears filament- or ridge-like at times. Elongated and filamentary-like structure is similarly seen, for example in the Chamaeleon-Musca filament \citep{Bonne2020} and in the diffuse ISM \citep[for a review see][]{Hacar2023}.

\begin{figure*}
    \centering
    \includegraphics[width=0.99\textwidth]{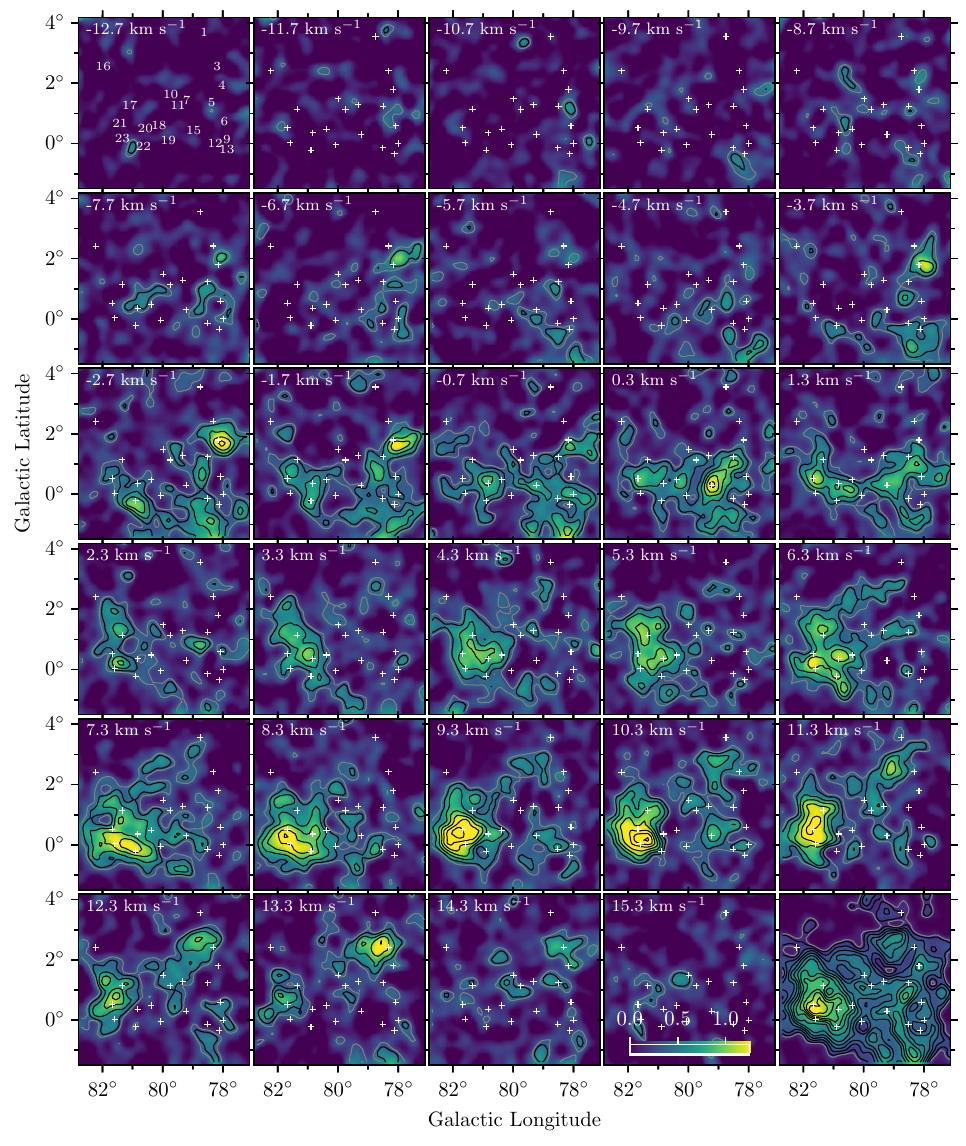}
    \caption{Channel maps of \refn{}. Each channel map has been integrated over 2 channels, equivalent to 1~\kms{}. The central velocity of the map is shown in its top left corner. In the first channel, numbers indicate the DR identifications \citep{Downes1966}; these sources are marked with a ``+'' in subsequent channel maps. The color bar, shown in the last channel, indicates the magnitude of $T_{\mathrm{C}273\alpha}/T_{\mathrm{C}}$ in units of 10$^{-3}$, and the color scale is the same for all maps. Gray contours are drawn at 3$\sigma$ and black contours are drawn at $[4, 6, 8, 10, 12, 14] \sigma$ where $\sigma_{\mathrm{median}} \approx 1.9 \times 10^{-4}$. The bottom right panel shows the \refn{}  emission integrated over $-$13 to 17~\kms{}; black contours are $[4, 6, 8,... , 24] \sigma$.}
    \label{fig:chan_CRRL}
\end{figure*}

Bright \refn{} emission peaks close to DR4, the southern edge of the supernova remnant (SNR) $\gamma$-Cygni, from channels -3.7~\kms{} to -1.7~\kms{}. The SNR is likely interacting with the ISM \citep[e.g.,][]{Ladouceur2008}. \cite{Roshi2022} analyzed RRL emission at 321 MHz within a single GBT beam in this location and found relatively bright carbon RRL emission, being equal in peak intensity to that of hydrogen RRLs at the same frequency. They argued for the CRRLs being emitted in a cool ($T = 20-200$~K) and dense ($n_e = 1.4 - 6.5$~\cmc{}) layer, likely compressed by a shock. Bright CRRL emission also appears towards DR3, the northern edge of the SNR, at 13.3~\kms{}. We discuss emission surrounding the $\gamma$-Cygni SNR in more detail in Section~\ref{ssec:discuss_DR4}.

%%%%%%%%%%%%%%%%%%%%
\subsection{\refn{} Spectra and Line Fits}

Figure~\ref{fig:spec_spatialavg} shows line emission from multiple tracers averaged over the full area of the survey region. In this Figure, \refn{} is presented in terms of a line brightness in units of K, the only instance where we do not show it in terms of $T_L/T_C$. Most emission between about -20 and 20 \kms{} in Figure~\ref{fig:spec_spatialavg} is attributed to clouds in the Cygnus X region forming a coherent complex. Only some velocity ranges can be attributed to emission from the Cygnus  rift at distances $< 1$~kpc. Emission at -40 \kms{} is from the Perseus Arm much further away. 

\begin{figure}
    \includegraphics[width=0.45\textwidth]{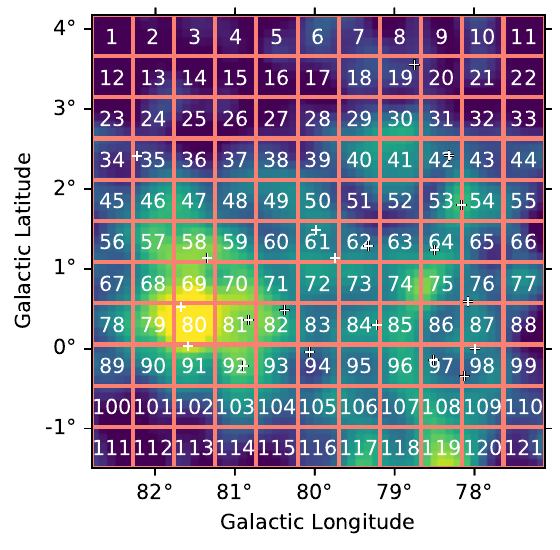}
    \caption{Aperture locations (salmon colored grid lines) and IDs (white number) for the extracted spectra shown in Figure~\ref{fig:specgrid_multiphase}.}
\label{fig:spec_apertures}
\end{figure}

\begin{figure*}
    \includegraphics[width=\textwidth]{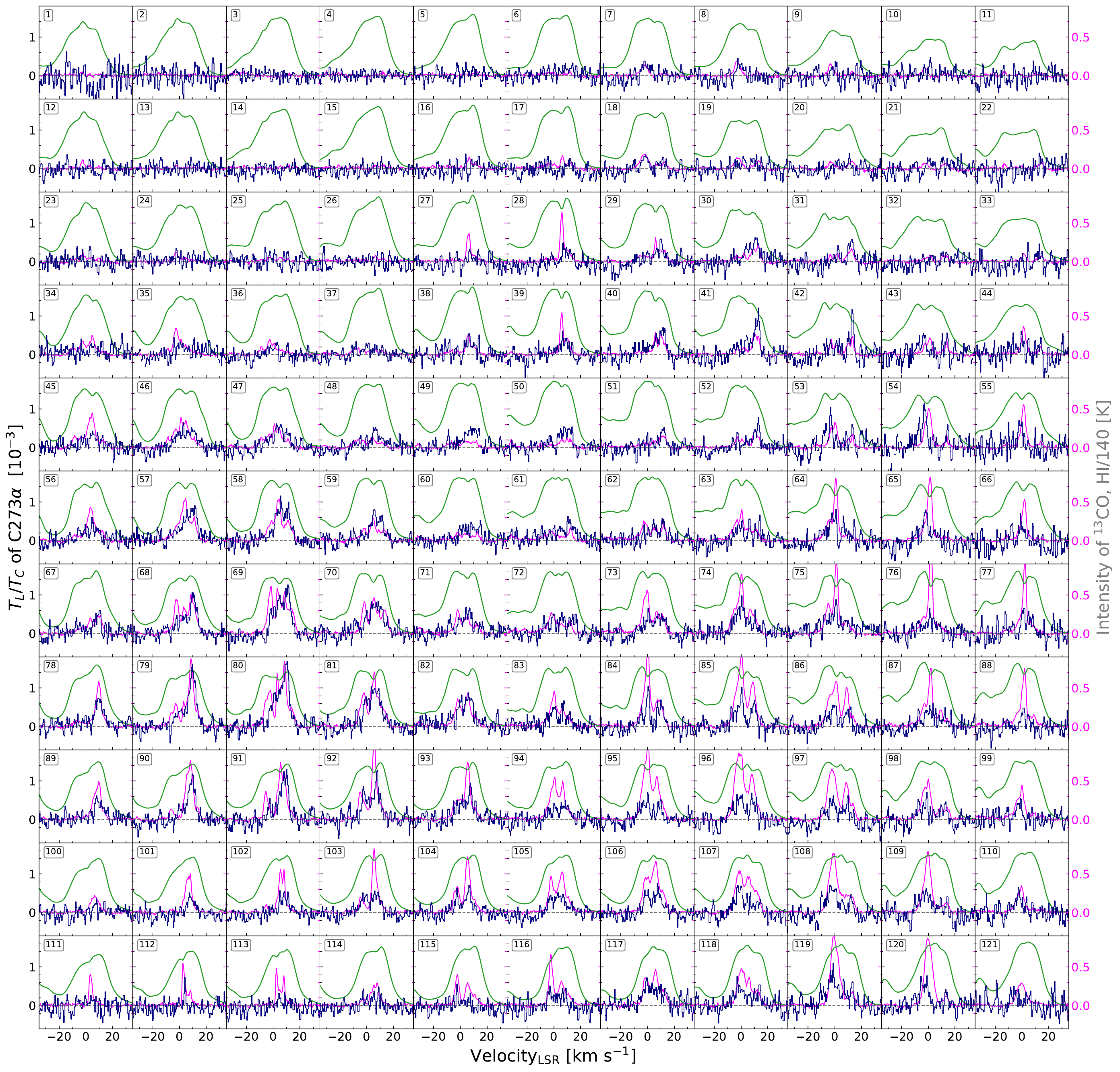}
    \caption{Multi-phase spectra over the mapped region. \refn{} emission is dark blue, \treceCO{} emission is in magenta, and \Hi{} 21 cm is green. The aperture ID is shown in a box in the upper-left corner of each spectrum. The y-axis on the right indicates the intensity in units of K for the \treceCO{} spectra and for  \Hi{} spectra that have been divided by a factor of 140.}
    \label{fig:specgrid_multiphase}
\end{figure*}

We also extracted \refn{} spectra from square apertures with a size of 5 pixels, or equivalently 30.9\amin{}, on a side. We show the aperture locations and IDs in Figure~\ref{fig:spec_apertures}. The spectra are shown in Figure~\ref{fig:specgrid_multiphase}. Overlaid on each spectrum are \Hi{} 21 cm and \treceCO{} spectra that have been extracted in the same apertures from data at matched resolutions (48.5\amin{} and 0.5~\kms{}) and voxel grid as \refn{}. 

The \refn{} emission is present in a majority of the aperture spectra. The \refn{} line profiles are Gaussian-like, indicating Doppler broadening by thermal, turbulent, or multiple velocity components. The profiles do not show signs of Lorentzian profiles with broad wings that have been observed in CRRLs, typically at lower frequencies ($\lesssim 100$~MHz), due to radiation or pressure broadening \citep[e.g.,]{Salgado2017b, Salas2017}. Doppler-broadened profiles are consistent with other P-band (300--400 MHz) observations of CRRLs \citep{Kantharia1998a, Roshi2000, Oonk2017}. 

When \treceCO{} emission appears, \refn{} is typically bright enough to be detected. However intensity ratios of the \refn{} and \treceCO{} do change by factors of more than 3. Interestingly, offsets in the central velocities of the \treceCO{} and \refn{} emission are apparent, for example see aperture IDs 53, 68, and 79 to name a few. We quantify \treceCO{} and \refn{} velocity differences and intensity ratios in Section~\ref{sec:compare_CO}.

\Hi{} emission is present in all apertures and has a fairly consistent intensity, unlike \refn{}. The intensity ratio of the \refn{} and \Hi{} changes by a factor of more than 10 across the region. In a number of apertures, a local dip in the \Hi{} spectrum coincides with a peak in \refn{} and/or \treceCO{} emission --- for example, apertures 28 and 84 --- and which may likely be \Hi{} absorption. However, the coincidence of \Hi{} absorption and \refn{} or \treceCO{} emission is not a consistent phenomenon.

We fit Gaussian profiles to the \refn{} spectra. To identify significant emission for fitting, we smoothed the data cube to 2~\kms{}, identified channels with a S/N~$>3$, and used the number of peaks in a contiguous chunk of channels as input for the number of components to fit. The fits were performed on the 0.5~\kms{} channel resolution data. We report components which have a $>5\sigma$ Gaussian area of the fit or a velocity-integrated intensity within the full width half-maximum (FWHM) of the fit. In total 122 components are reported from the \refn{} emission. Spectra showing the best-fit profiles and residual spectra are shown in Appendix Figure~\ref{ap:fig:crrl_fits}.

\begin{figure}
    \flushright
    \includegraphics{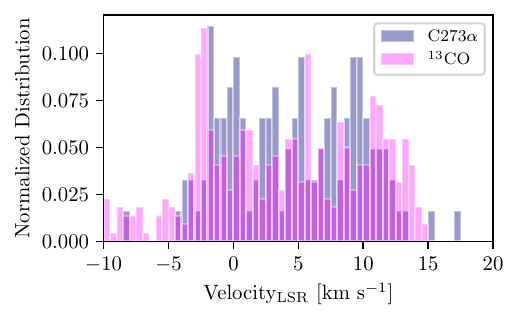}
    \includegraphics{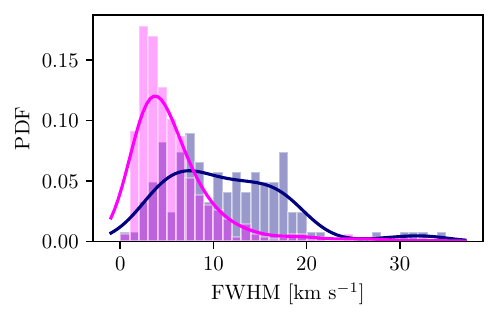}
    \includegraphics{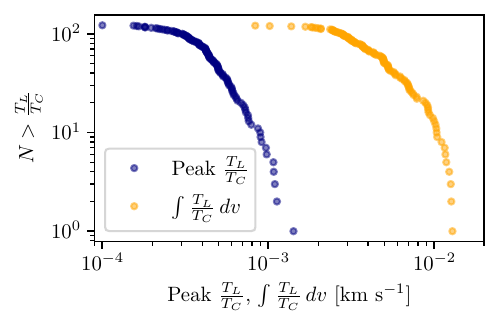}
    \caption{Best fit parameters of Gaussian profiles of \refn{} (dark blue) and \treceCO{} (magenta) across the mapped region. \textit{Top} is distribution of the best-fit central velocities. \textit{Middle} is the distribution of best-fit FWHM. \textit{Bottom} is the inverse cumulative sum of the best-fit amplitudes (dark blue) and integrated line strength (orange).}
    \label{fig:line_props}
\end{figure}

Properties of the fitted line profiles are shown in Figure~\ref{fig:line_props}. The central velocities span -9~\kms{} to 17~\kms{}. The amplitudes of the line fits have a median of $4 \times 10^{-4}$, with the brightest component having a line-to-continuum ratio of $1.4 \times 10^{-3}$. The distribution is steep, rapidly increasing in number towards lower amplitude values. 

The line widths span a large range, with FWHM from 2 to 20~\kms{}. The median is FWHM$_{\mathrm{C}273\alpha} = 10.6$~\kms{} (corresponding to a velocity dispersion of $\sigma_{\mathrm{C}273\alpha} = 4.5$~\kms{}) and with a typical uncertainty of 1.1~\kms{}. The line-width distribution (Figure~\ref{fig:line_props}) may truly be bimodal; as we plot the distribution of ever higher S/N features, the distribution skews towards larger line widths.  We caution the reader of the low S/N of these data. Higher sensitivity observations would be very useful to assess the line width distribution with greater certainty. 

The \refn{} line widths are  considerably larger than purely thermal broadening ($= \sqrt{8 \ln{2}} \cdot \sqrt{2kT/m}$) of 0.5--1~\kms{}, of C$^+$ ions at temperatures of 20--100 K.
The typical velocity difference found for a \refn{} component with respect to \treceCO{} is 2.9~\kms{} (see Section~\ref{sssec:vel_offset}). Thus we might expect a single component to contribute line broadening on the order of 3~\kms{}, however the majority of the line widths are broader than this. {Unresolved cloud components or turbulent motions may therefore dominate the line broadening of the \refn{}.

Here we consider the scenario in which the \refn{} line widths are dominated by turbulent motions. The isothermal sound speed of \Hi{} gas, adopting a mean atomic weight of 1.36$m_H$ so that $C_s^2 = (k_B T / 1.36m_H)$, is $C_s = 0.78$~\kms{} at a temperature of 100~K. The isothermal sound speed of \Htwo{} gas, adopting a mean atomic weight of 2.36$m_H$ so that $C_s^2 = (k_B T / 2.36m_H)$, is $C_s = 0.26$~\kms{} at a temperature of 20 K. This implies that the median 1D velocity dispersion of 4.5~\kms{}, equal to a 3D velocity dispersion of $\sigma_{3D = \sqrt{3} \sigma_{v,1D}} = 7.8$~\kms{}, implies Mach numbers, $\mathcal{M} = \sigma_{3D} / C_s$, somewhere between 10--30. 
These Mach numbers are rather high with respect to values obtained for CNM gas in diffuse ISM conditions, $1 < \mathcal{M} < 4$ \citep{Heiles2003b, Jenkins2011}. Higher turbulent pressures and line widths are typically found in a region with high star formation activity as compared with diffuse ISM clouds. A range between $10 < \mathcal{M}_{CRRL} < 30$ should be considered as an upper limit to the representative Mach number of the \refn{} gas, since an observed narrower velocity dispersion would translate to smaller Mach numbers. 

We also point out that the large spread in line widths might also imply gas that can be found in a variety of states. The low-end dispersion of 2~\kms{} implies some gas is present with Mach numbers of 2--7, more typical of the diffuse ISM. At the high-end, the dispersion of 9~\kms{} becomes less certain (due to the possibility of contamination with multiple velocity components), but would imply Mach numbers of 20--60. Deeper observations of \refn{} emission would be useful and necessary to measure its intrinsic unbiased line width on these spatial scales.

%%%%%%%%%%%%%%%%%%%%%%%%%%%%
\section{Comparison with \treceCO{}}
\label{sec:compare_CO}

In this section we compare \treceCO{} and \refn{} emission. We use \treceCO{} data cubes that are matched spatially and spectrally in resolution and grid with the \refn{} data cubes.

%%%%%%%%%%%%%%%%%%%%%%%%%%%%%%%%%%%%
\subsection{$^{13}$\text{CO} Channel Maps}

%%%%%%%%%%%%%%%
\begin{figure*}
    \centering
    \includegraphics{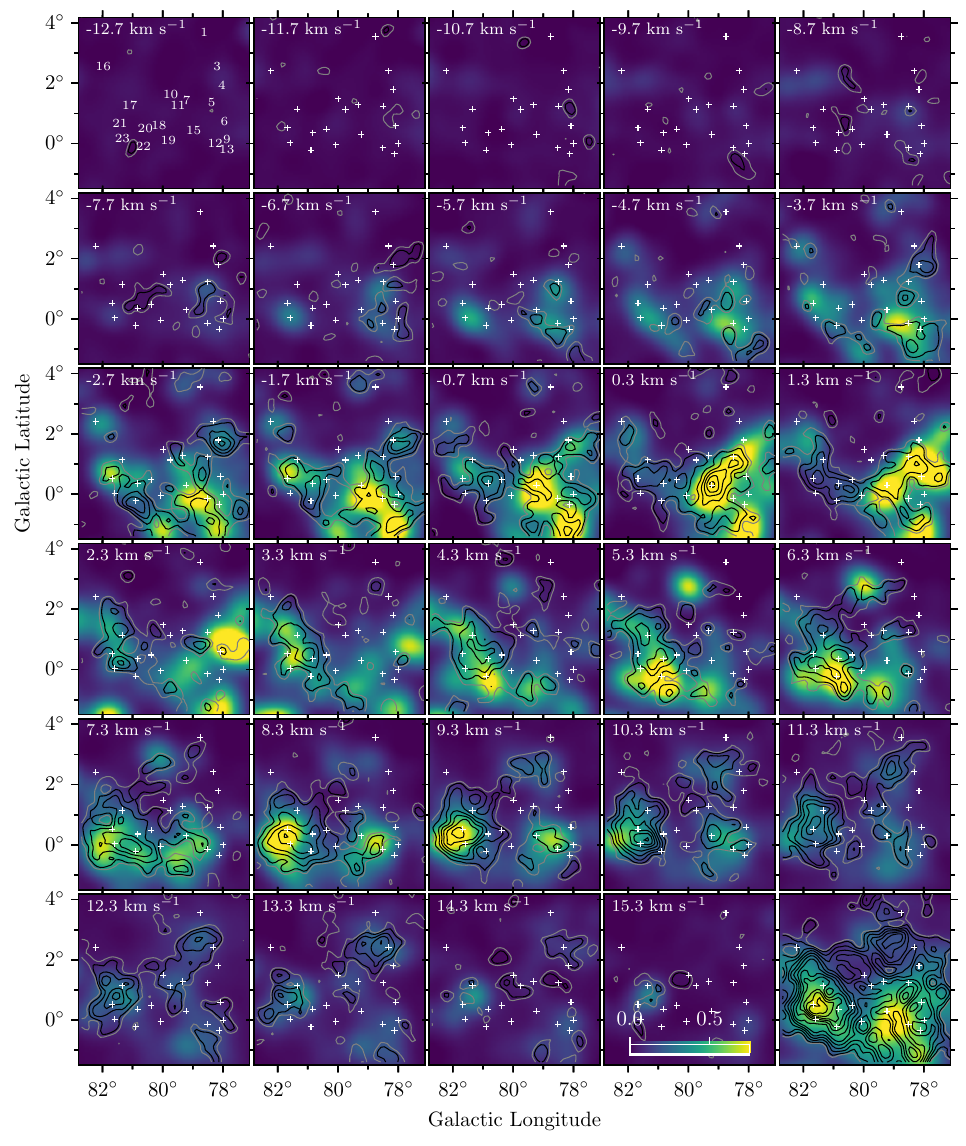}
    \caption{\treceCO{} intensity, in units of K (see colorbar in Channel 16.3~\kms{}), at matched resolutions to \refn{}.  \refn{} contours, DR IDs, and ``+'' are the same as in Figure~\ref{fig:chan_CRRL}. The bottom right panel shows the integrated intensity of \treceCO{} over -13 to 17~\kms{}, with \refn{} contours overlaid.} 
    \label{fig:chanmaps_CO}
\end{figure*}

Channel maps of the MWISP \treceCO{} emission \citep{Su2019} with \refn{} contours overlaid are shown in Figure~\ref{fig:chanmaps_CO}.  Overall the \refn{} emission appears to coincide with the velocities where \treceCO{} emission is present. However, their morphologies are noticeably different.  \refn{} appears where \treceCO{} is both relatively faint and bright. Peaks in emission are often spatially offset, and \refn{} is more often than not found on the outskirts of \treceCO{} clouds.
There are numerous examples of \refn{} spatially offset from \treceCO{} ridges, for example at -2.7~\kms{} below DR 15/12/13 and below DR 23, most of the emission in Channel 3.3~\kms{} that forms a ridge, and in Channel 1.3~\kms{} just below DR 13 there is a distinct offset alongside CO emission. 

There are also interesting regions where CRRLs are detected with strong significance but \treceCO{} is comparatively fainter. This occurs generally in velocity channels of +10.3~\kms{} and higher. \refn{} around DR 18/19/20/22 bridges two \treceCO{} clouds in Channel 0.3--1.3 ~\kms{}. \refn{} emission in Channel -0.7~\kms{} hugs the \treceCO{} clouds below DR 22 and DR 23. Emission just above and to higher longitudes of DR 10 appears to ``connect'' two \treceCO{} clouds starting at 6.3~\kms{} and continuing to 13.3~\kms{}.
The linear extent of \refn{} emission in Channel 7.3~\kms{} is not matched in \treceCO{} emission.

At this matched resolution, corresponding to about 21 pc, \refn{} shows considerably more structure than \treceCO{}. This might arise from and possibly indicate the ubiquity of CO emission on numerous scales, whereas \refn{} emission may be less volume filling and/or more narrow along some dimensions.

%%%%%%%%%%%%%%%%%%%%%%%%%%%%
\subsection{Velocity-Integrated Intensity Comparison}

Figure~\ref{fig:co-crrl-intens} compares the \refn{} and \treceCO{} emission, showing the velocity-integrated emission of each 2D spatial pixel. Included are data points where \refn{} has a S/N$\geq 3$, for which all \treceCO{} points have a S/N$>5$. An overlay of the Moment 0 maps used for this analysis can be found in Figure~\ref{fig:chanmaps_CO} and in the Appendix Figure~\ref{ap:fig:overlay_matched_res}. At the present $\sim$21~pc resolution, \refn{} emission from cool dark gas is present over a sizable range of \treceCO{} integrated intensity and in proportion, column density.

There may be a mildly increasing positive trend, as the intensity of \treceCO{} increases so does \refn{}, and we compute a Pearson correlation coefficient of $r = 0.63$.  The power-law fit, $\log_{10} y = a \log_{10} x + b$, we find using the ordinary least-squares bisector method is $a = 0.46 \pm 0.07$ and $b = -2.50 \pm 0.05$.

\begin{figure}[t]
    \centering
    \includegraphics[width=0.495\textwidth]{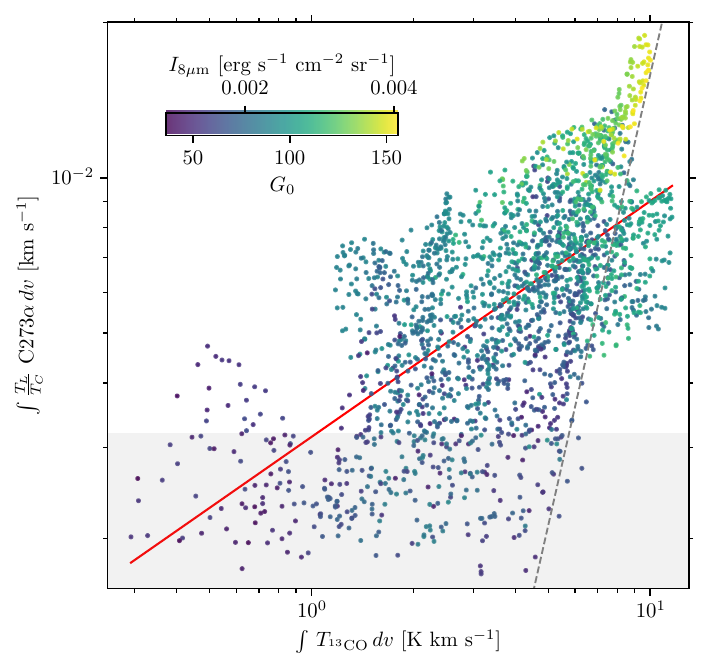}
    \caption{A comparison of the point-by-point velocity-integrated \treceCO{} and \refn{} emissions. The gray shaded region indicates where \refn{} has $3\leq S/N <5$. We note all \treceCO{} points have a S/N$>5$. The data points are colored by $I_{8~\mu m}$ and the corresponding estimate of $G_0$ (see Section~\ref{sec:8micron}). The red solid line shows the best-fit power-law with slope $a=0.46$, while the gray dashed line has a power-law slope of $a=2.9$ and has been included for illustrative purposes of a possible high intensity regime.}
    \label{fig:co-crrl-intens}
\end{figure}

However, the distribution has noticeable systematic deviation and gaps about the trend. For example, the brightest \refn{} emission protrudes up away from the best-fit trend. These pixels are also associated with the brightest 8 $\mu$m intensity, as indicated by the color of the data points in Figure~\ref{fig:co-crrl-intens} and correspond to $G_0 \gtrsim 140$ (see also Section~\ref{sec:8micron}). It may reflect a higher intensity regime somewhat correlated with massive star formation and column density. We do note caution in over-interpreting the data here, as the pixels sample approximately one beam. Follow up investigations of other high intensity regimes and/or at higher spatial resolution would be highly valuable to better understand the interdependence of the cool dark gas (\refn{}), CO-traced molecular gas (\treceCO{}), and the radiation field (8~$\mu$m).

On one hand, a mild correlation between \refn{} and low-intensity \treceCO{} can be expected. The low-frequency CRRLs are expected to arise in an \Av{} (or likewise, column density) that is (weakly) dependent upon $G_0/n$ \citep{Wolfire2010, Wolfire2022}. At lower $G_0/n$ the \Hi{}-to-\Htwo{} transition (and the C$^+$/C/CO transition), moves to lower \Av{}. The \treceCO{} intensity is directly proportional to column density (\Av{}). On the other hand, CO cloud-layers are not particularly sensitive to the strength of the radiation field (see also Appendix Figure~\ref{ap:fig:more_fluxcomp}), as the deeper cloud layers are sufficiently shielded. With the present spatial resolution, a large range of \treceCO{} clouds and varying properties fall within one resolving beam, including high column density regimes that are well separated from a C$^{+}$ layer, and result in decorrelation of the tracers.

Lastly we note that variation in cloud distances, spanning up to $\approx$1.3 -- 1.7~kpc, could account for some variation, up to a factor of 1.7, in the observed trends. The CRRL emission is dependent upon the background continuum intensity and \textit{independent of the distance} along the line of sight to the background continuum source (see Section~\ref{sec:crrl_emission}), whereas the beam-averaged \treceCO{} intensity may indeed vary with the distance to its emitting cloud. There is a known molecular cloud, the Great Cygnus Rift, in this direction at a distance of approximately 600 -- 800 pc. However, the correlation of \refn{} with 8~$\mu$m (see Section~\ref{sec:8micron}) suggests that the majority of CRRLs are related to the primary source of the FUV radiation in this direction, that is, associated with the Cygnus X region. The Cygnus X region is thought to be a coherent region of active star formation at about 1.5~kpc \citep[e.g.,][]{Schneider2006}, with differences in the cloud distances not exceeding $\pm 0.2$~kpc \citep[e.g.,][]{Rygl2012, Quintana2021}.

%%%%%%%%%%%%%%%%%%%%%%%%%%%%%%%%%%%%%%%%%%%%%%%%%%%%%%%%%%%%%
\subsection{Spatial Separation of \refn{} and \treceCO{} emission}
\label{ssec:spatial_sep}

We compare the angular separation of peaks in \refn{} and \treceCO{} emission in each channel map. We identified emission peaks through dendrogram structures, making use of the Python package, \texttt{Astrodendro} \citep{Rosolowsky2008}. Dendrograms locate islands of pixels with increasing-only intensities around a local maximum in emission. For \refn{}, we used a threshold of 5$\sigma$ and a minimum of three pixels to define peaks. For \treceCO{} emission, we set the threshold of 0.1 K~\kms{} and a minimum of three pixels above the threshold.

For each peak of \refn{} emission, we located the \treceCO{} peak in the same channel which was closest in angular separation. We plot the probability density function of these results in Figure~\ref{fig:spatial_sep}, which was estimated using a Gaussian kernel and applying Scott's Rule to determine the bin size. The distribution shows a peak at 26\amin{}, corresponding to  12 pc. 26\amin{} is close to half the size of the beam (24\amin{}), but it is much larger than the pointing accuracy of the observations (1\amin{}). 
The distribution of angular separations between \refn{} and \treceCO{} peaks quantifies the differences that can be seen by-eye in Figure~\ref{fig:chanmaps_CO}.

%%%%%%%%%%%%%%
\begin{figure}
    \centering
    \includegraphics[width=0.48\textwidth]{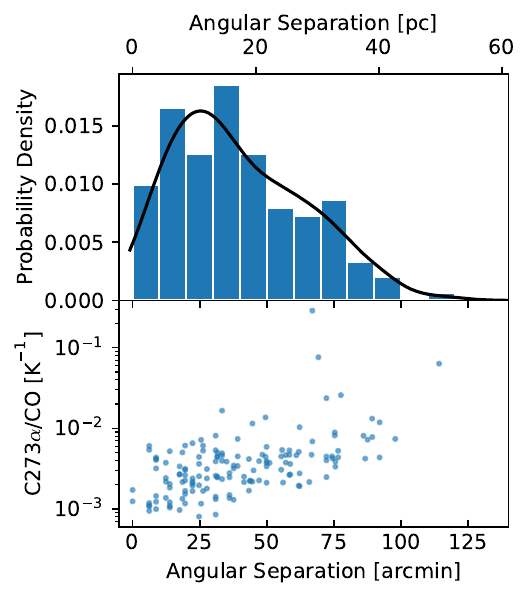}
    \caption{\textit{Top:} Spatial separation between \refn{} peaks and the closest \treceCO{} peak identified in channel maps. A maximum in the distribution is found at 26\amin{} (12 pc), which is larger than the pointing accuracy of 1\amin{}. \textit{Bottom:} The angular separation plotted against the \refn{}/$^{13}$CO intensity ratio at the pixel of the \refn{} peak. The intensity ratio of \refn{}/$^{13}$CO tends to be larger when the angular separation between intensity peaks is larger.}
    \label{fig:spatial_sep}
\end{figure}

%%%%%%%%%%%%%%%%%%%%%%%%%%%%%%%%%%
\subsection{Comparisons with \treceCO{} line profiles}
\label{ssec:kinematic_diff}

With the high signal-to-noise nature of \treceCO{} data, (unlike \refn{}), we were able to use \texttt{GaussPy+} to automate fitting the aperture spectra of Figure~\ref{fig:specgrid_multiphase}. Examples of the fitted components and residual spectra and a description of the fitting procedure in Appendix~\ref{ap:sec:linefitting}. In total, 496 components were fit to the \treceCO{} spectra. 

%%%%%%%%%%%%%%%%%%%%%%%%%%%%%%%%
\subsubsection{Velocity Offsets}
\label{sssec:vel_offset}

In Figure~\ref{fig:line_props} we show the central velocities of all fitted \treceCO{} components overlaid with \refn{} components. Curiously the central velocities of \refn{} and \treceCO{} components appear to be somewhat anti-correlated, where the \refn{} emission appears at local deficits of \treceCO{}. The typical error of the \treceCO{} velocity centers is 0.2~\kms{}.

We also compare the difference in central velocities between \treceCO{} and \refn{} components. For each fitted \refn{} component in an aperture, we identify the best-fit \treceCO{} component that falls closest in velocity, defined as having the smallest absolute value of the velocity difference. We plot these results in Figure~\ref{fig:vel_offset}. The PDF is Gaussian-like.  The standard deviation of the velocity differences, 2.9~\kms{}, is larger than the combined errors of the fitted centers, 0.53~\kms{}. This quantifies the velocity differences which can be seen by eye in the spectra and channel maps of the two tracers.

%%%%%%%%%%%%%%
\begin{figure}
    \centering
    \includegraphics[width=0.45\textwidth]{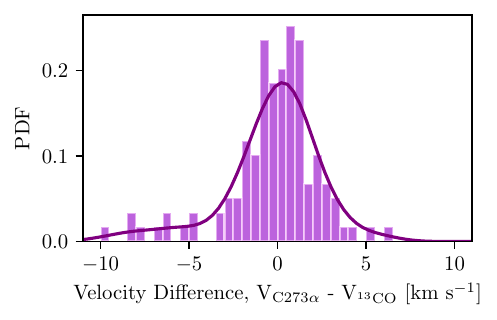}
    \caption{The velocity difference between each \refn{} component and the \CO{} component nearest in velocity for a given aperture. The standard deviation of this distribution is 2.9~\kms{}.}
    \label{fig:vel_offset}
\end{figure}

Furthermore, the center of the $\mathrm{V_{C273\alpha} - V_{^{13}CO}}$ distribution, 0.2~\kms{}, is consistent with zero within the error. 
When we sub-select for different regions or velocity groupings in the map, the $\mathrm{V_{C273\alpha} - V_{^{13}CO}}$ distribution is consistently centered about zero within error. This implies that the flow of the \refn{} gas is not dominated by one systematic velocity, but rather a distribution of both red- and blue-shifted velocities. Since the spatial resolution is limited and several velocity components are present per aperture, it is plausible that velocity differences determined at higher spatial resolution could show a larger magnitude of difference.

Numerical studies and observations are finding that dynamical effects may be an important aspect to the molecular formation process, rapidly speeding up the timescales over which \Hi{} is converted into \Htwo{} \citep{Glover2007b, Beuther2014, Valdivia2016, Gong2017, Bialy2017, Bisbas2017b, Clark2019b, Heyer2022}. 

\cite{Park2023} found velocity differences between cold \Hi{} and CO of 0.4~\kms{} towards an ensemble of local diffuse clouds, whereas velocity differences between warm \Hi{} and CO in their sample are 1.7~\kms{}. A smaller velocity difference in their diffuse clouds could be the result of different dynamics at play due to star formation and stellar feedback and/or the higher density environment that is found in Cygnus X. 

Velocities of cool neutral gas in the range 1-4~\kms{} are typically found for gas infalling under gravitational collapse on pc scales \citep{Schneider2010, Beuther2015, Dhabal2018, Williams2018, Wang2020, Bonne2020, Heyer2022, Bonne2023}. While these are similar to what we measure, in Section~\ref{ssec:region_props} we reason that the  between \refn{} and \treceCO{} velocity differences are not the result of gas under gravitation collapse.Conversely, with $10^6$ \Msun{} in a region with a diameter of $\sim$100 pc, the escape velocity is $\sim$10~\kms{}. We do not (yet) see clear evidence for this kind of coherent velocities that could indicate that material is blown away before it can participate in star formation.

%%%%%%%%%%%%%%%%%%%%%%%%%%%
\subsubsection{Line widths}

Figure~\ref{fig:line_props} also shows the distribution of line-widths of the fitted \treceCO{} components with the \refn{} components. A median value of FWHM$_{^{13}\mathrm{CO}} = 2.5$~\kms{} and error of 0.5~\kms{} is found for \treceCO{}.
The \refn{} line width of FWHM$_{\mathrm{C}273\alpha} = 10.6$~\kms{} is comparatively larger than that found for \treceCO{}. The distribution of line widths is significantly more peaked for \treceCO{}. In Figure~\ref{fig:spec_apertures}, the \treceCO{} and \refn{} show emission over the same velocity ranges but the higher S/N of the \treceCO{} data breaks the emission profile up in multiple components, each with a width considerably less than the overall profile. The limited S/N of the \refn{} hampers the separation in components but it is likely that multiple components are present and the line width of such components might well be similar to those of \treceCO{}. Confirmation awaits higher S/N data.

{We may also consider that the broad distribution of the \refn{} line widths is accurately representative. While \treceCO{} may consist of both dense filament like structure as well as fluffy diffuse components, it does not show a bimodal or broad distribution of line width. \refn{} may be dynamically more active and variant that \treceCO{}. The differences in the line width distributions could also be influenced by the \refn{} emission being weighted by density squared (i.e., $EM$), rather than density as for \treceCO{}. Whereas a diffuse \treceCO{} cloud component can dominate over narrow, dense filaments at larger scales, the same may not be the case for CRRL emission. This is supported by \refn{} showing more variation in emission structure, resulting in a less cloud-like and more ridge-like morphology. 

The spread in the \treceCO{} line width distribution is significantly smaller than for \refn{}, and is well represented by its characteristic value. Mach numbers of 12--14 are derived in gas with temperatures of 15--20~K, consistent  with $5 < \mathcal{M} < 20$ typically found within CO-emitting molecular clouds \citep{Zuckerman1974, Brunt2010}. It is interesting that at least some of the \refn{} emission has similar Mach numbers as the \treceCO{} material, which would be expected for gas tracing similar (i.e., \Htwo{}) states.

%%%%%%%%%%%%%%%%%%%%
\section{Comparison with 8~$\mu$\MakeLowercase{m} emission} 
\label{sec:8micron}

For this analysis, we convolved and re-gridded the MSX 8~$\mu$m image to the \refn{} beam size and pixel size.  An overlay of the velocity-integrated emission maps of the two tracers can be found in Figure~\ref{ap:fig:overlay_matched_res} of the Appendix. Figure~\ref{fig:8micron} plots the velocity-integrated \refn{} emission against the 8~$\mu$m intensity point-by-point. A good correlation is found which has a Pearson correlation coefficient of $r = 0.76$. The power-law fit, $\log_{10} y = a \log_{10} x + b$, we find using the ordinary least-squares bisector method on \refn{} pixels $\geq 5 \sigma$ is $a = 1.3 \pm 0.2$ and $b = 1.3 \pm 0.5$.

We estimate the FUV radiation field in units of $G_0$ using the 8 $\mu$m PAH intensity. We apply the relations of 8 $\mu$m intensity with FIR intensity and $G_0$ determined by \cite{Pabst2021, Pabst2022}, using $G_0 \simeq I_{8 \mu m} / 2.6 \times 10^{-5}$~erg~s$^{-1}$~cm$^{-2}$~sr$^{-1}$. $G_0$ estimates, $\approx 40 - 165$ with a median of 90, are dominated by the diffuse large-scale structures in the region (e.g., see Figure~\ref{fig:cyg_intro}) and are elevated above typical ISM values. The $G_0$ values relatively agree with those from the FIR intensity that has been determined for a subsection of the region \citep{Schneider2016b}.

Furthermore, the dependence of the \refn{} emission on the FUV radiation suggests that the CRRLs are indeed dependent upon the FUV field which predominantly originate from the Cyg X region instead of foreground emission along the line of sight. It is not expected that integrated interstellar radiation field along the line of sight would appreciably contribute and result in a systematic offset of the reported $G_0$ values.

We interpret the \refn{} and 8~$\mu$m correlation as the dependence of the cool dark gas emission on the FUV radiation field, which can be expected for diffuse PDRs as summarized by \citet{Hollenbach1999} \citep[see also][]{Wolfire2022}. The power-law exponent of $\simeq 1.3$ may provide insight into the heating/cooling efficiency of the gas, as C$^+$ dominates cooling of this gas phase \citep{Tielens2005}. We note the agreement of the slope measured for low-intensity [CII] ($I_{\mathrm{[CII]}} \lesssim 4 \times 10^{-2}$~erg~s$^{-1}$~cm$^{-2}$) in Orion A of $a = 1.2 \pm 0.8$, which is notably steeper than the slope of $a = 0.6 \pm 0.3$ observed for their higher-intensity [CII] \citep{Pabst2021}.

In Figure~\ref{fig:8micron} we color the data points depending on their velocity-integrated \treceCO{} intensity, finding no obvious trends, but instead a bit of a mix of \treceCO{} intensities throughout, which is not unexpected for nonhomogeneous molecular cloud properties throughout the region.

%%%%%%%%%%%%%%
\begin{figure}
    \centering
    \includegraphics[width=0.47\textwidth]{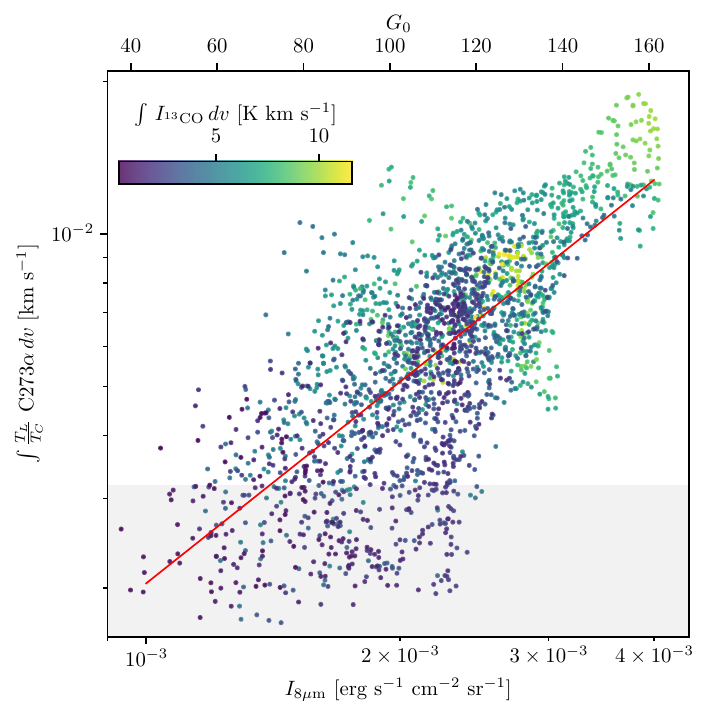}
    \caption{A comparison of the point-by-point velocity-integrated \refn{} and the 8~$\mu$m emissions, revealing a correlation. The top axis shows the corresponding FUV radiation $G_0$ estimated from the 8~$\mu$m intensity. The gray shaded region indicates where \refn{} has $3\leq S/N <5$, and the data points are colored by $\int I_{^{13} \mathrm{CO}} \,dv$. The red solid line shows the best-fit power-law with slope $a=1.3 \pm 0.2$.}
    \label{fig:8micron}
\end{figure}
%%%%%%%%%%%%

%%%%%%%%%%%%%%%%%%%%
\section{Discussion} 
\label{sec:discuss}

%%%%%%%%%%%%%%%%%%%%%%%%%%%%%%%%%%%%%%%%%%%%%%%%%%%%%%%%%%%%%
\subsection{Spatial dependencies on the intensity relations?}

\begin{figure*}[t]
    \centering
    \includegraphics[width=0.32\linewidth]{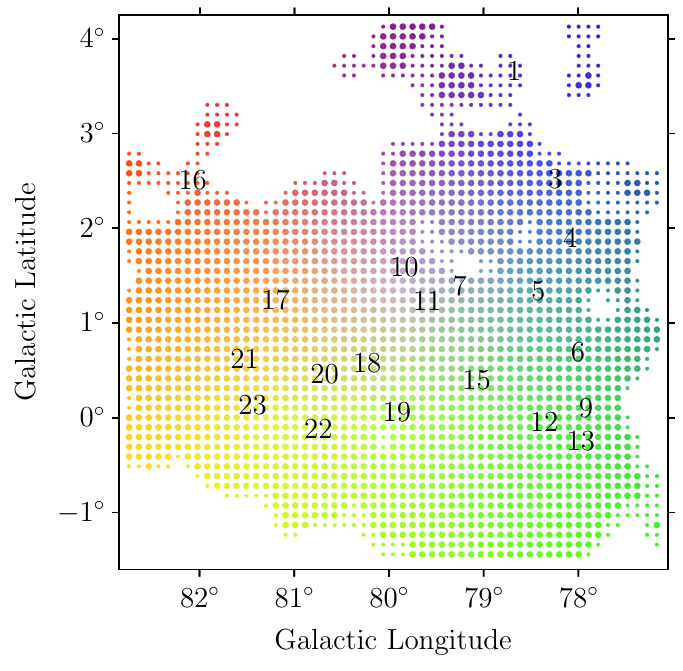}
    \includegraphics[width=0.66\linewidth]{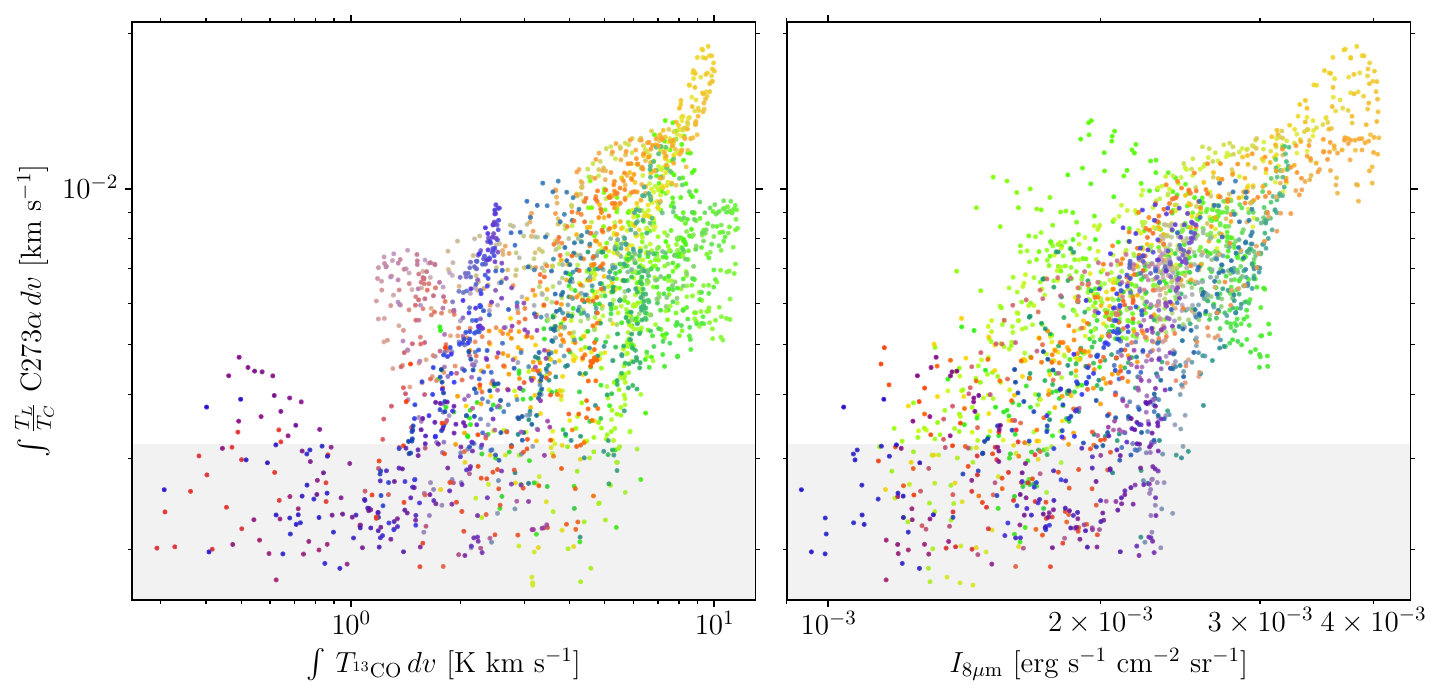}
    \caption{The 2D colormap \textit{(left)} of \refn{} with S/N$\geq5$ (big data points) and $3 \leq S/N < 5$ (small data points) is used to discern if/how the spatial location in the region maps to the intensity comparisons of \treceCO{} \textit{(middle)} and 8~$\mu$m \textit{(right)}. Black numbers indicate the DR region ID, same as in Figure~\ref{fig:cyg_intro}. Co-spatial regions tend to group together in the \treceCO{} plot but not in 8~$\mu$m emission. }
    \label{fig:fluxcomp_colorspatial}
\end{figure*}

We use a 2D color map (Figure~\ref{fig:fluxcomp_colorspatial}, \textit{left}) to investigate how spatial locations map in the intensity relations presented in Figures~\ref{fig:co-crrl-intens} and \ref{fig:8micron}. Plots of the intensity relations with data points colored by their 2D spatial location are shown in Figure~\ref{fig:fluxcomp_colorspatial}. For the 8~$\mu$m comparison, no trends emerge. However, with the \treceCO{} data, co-spatial regions do tend to fall in similar locations on the plot, as can also been seen in the comparison of 8~$\mu$m and \treceCO{} in Appendix Figure~\ref{ap:fig:more_fluxcomp}. This supports the interpretation that differences in local conditions, namely average density of the cloud, are influencing its average location in the plot, both with respect to \treceCO{} and with respect to the distribution of \refn{} emission which may depend on both density squared and temperature.

For some of the co-spatial (i.e., same color) data points, the slope of \treceCO{} and \refn{} seems to be fairly similar, but with a normalization that varies. In these regions, a modest change in \treceCO{} corresponds with a large increase in \refn{}, and the maximum \refn{} is seemingly dictated by $G_0$. The deep blue streak corresponds to DR3 and the region surrounding NGC~6910; the yellow-mustard streak with DR21 and DR23; the orange streak with DR17; the dark cyan streak with DR5, a PDR possibly powered by NGC 6913; and, the dark green with DR6. In contrast, the light-green data points, corresponding to the massive CO clouds of Cyg X South (south of DR12 and DR13), show a smaller relative change in \refn{} and \treceCO{}. These effects may result from the (low spatial) resolution, different cloud properties, the Cyg X South clouds being slightly closer than the Cyg X North clouds and/or influencing their proximity to Cyg OB2. Higher spatial resolution and multi-band observations to measure the low-frequency CRRL SLED would provide understanding.

%%%%%%%%%%%%%%%%%%%%%%%%%%%%%%%%%%%%%%%%%%%%
\subsection{Estimated \refn{} gas densities} 
\label{ssec:discuss_spatialsep}

Steady-state analytical models have been constructed and widely applied to describe the \Hi{}-to-\Htwo{} transition in the ISM of galaxies \citep[e.g.,][]{Krumholz2008, Krumholz2009, McKee2010, Wolfire2010, Sternberg2014, Gong2017}.
Here we apply the framework of \citet{Wolfire2010} which focuses on cool dark gas in the \Hi{}-to-\Htwo{} and C/C$^+$ transitions at the edges of molecular clouds.

Figure~\ref{fig:wolfire_params} illustrates a model cloud and some of its defining parameters as laid out by \citet{Wolfire2010}. We depict what is thought to be the most likely scenario of the CRRLs: arising from a cold dense state of C$^+$ which immediately proceeds the C layer, and where the gas is mainly \Htwo{}. The \citet{Wolfire2010} model describes clumps of molecular clouds, with respect to $R_{\mathrm{CO}}$, the radius at which the optical depth of \CO{} as measured from the outer radius inwards is equal to 1, and the corresponding extinction depth into the cloud $A_V(\mathrm{CO})$; the radius, $R_{\mathrm{H}_2}$, and the extinction depth into the cloud, $A_V(\mathrm{H}_2)$ at which there is equal mass density in H atoms and \Htwo{} molecules; and, $R_{\mathrm{T}}$, the total cloud radius which includes (cold) \Hi{}. \citet{Wolfire2010} describe $A_V(\mathrm{H}_2)$ and $A_V(\mathrm{CO})$ in terms of key parameters, $n$, the local number density of the clump, $G_0$, the incident FUV radiation field, and $Z$ the metallicity. At a fixed metallicity, the ratio $G_0/n$ regulates the relation.

The ratios of the radii of these layers are found by \citet{Wolfire2010} to be robust, changing little under the conditions present, with nominal values of $R_{\mathrm{H}_2} \approx 1.2 R_{\mathrm{CO}}$ and $R_{\mathrm{T}} \approx (1.3 -1.4) R_{\mathrm{CO}}$.  We consider the most likely case in which the \refn{} arises from a C$^{+}$/\Htwo{} layer that falls within $R_{\mathrm{CO}}$ and $R_{\mathrm{H_2}}$, such that $R_{\mathrm{CO}} < R_{\mathrm{CRRL}} \leq R_{\mathrm{H_2}}$.

We use the angular separation derived in Sec.~\ref{ssec:spatial_sep} to represent $R_{\mathrm{CRRL}}$; this assumes that the center of the cloud is traced by the \treceCO{} peak, and so the computed separation between a peak in \treceCO{} and a peak in CRRL emission corresponds to the radius of the cloud's CRRL layer. Since a relation for $R_\mathrm{CRRL}$ and the fraction of $\Delta R(\mathrm{dark})$ it occupies is unknown,  we consider a range in cloud properties that are bounded by (a) the upper limit of $R_{\mathrm{CRRL}} =  R_{\mathrm{H}_2}$ and (b) the lower limit of $R_{\mathrm{CRRL}} = R_{\mathrm{CO}}$. We then use the velocity-integrated \treceCO{} measurements to ultimately derive an estimate of the volume density of the clump. We compute results using the median value of $\int T_{^{13}\mathrm{CO}} dv = 5$~K~\kms{}, as well as the approximate maximum (minimum) value of $\int T_{^{13}\mathrm{CO}} dv = 10 \,(1)$~K~\kms{}. We estimate a beam volume filling factor from the assumed $R_{\mathrm{CO}}$ of a spherical clump. Following \citet{Schneider2010}, we derive an \Htwo{} column density. We compute the \treceCO{} column density as $N_{^{13}\mathrm{CO}} = 1.1 \times 10^{15} \int T_{^{13}\mathrm{CO}} dv$, assuming a \treceCO{} excitation temperature of 15 K, the average value determined for the Cyg X North and South regions \cite{Schneider2006}. Then the \Htwo{} column density is computed as $N_{\mathrm{H}_2} = 4.7 \times 10^5 N_{^{13}\mathrm{CO}}$.

Table~\ref{tab:physprops} lists values which the model is evaluated at and the computed \Htwo{} column densities. Since the \Htwo{} column densities are relatively large, we make the assumption that the total cloud column density may be approximated as $N_{\mathrm{H}} \approx 2 N_{\mathrm{H}_2}$.

We estimate the H nucleus number density as $n = N_{\mathrm{H}} / (\frac{4}{3} R_{\mathrm{CO}}) / 1.2$, where the factor 1.2 approximates the difference between the local number density at $R_{\mathrm{CO}}$ and the input needed for the model, at $R_{\mathrm{H}_2}$. The estimated number densities are listed in Table~\ref{tab:physprops}. Using the median values for integrated \treceCO{} and angular separation, densities of $n = 300 - 500$~\cmc{} are found. We consider these to be representative of the densities at which the majority of \refn{} may arise. However, when considering the full range of integrated \treceCO{} intensities and large angular separation found when \treceCO{} is faint, the data indicate that \refn{} densities may fully encompass $n = 20 - 900$~\cmc{}.

We compute $A_V$ predicted by the \citet{Wolfire2010} framework using $n$ and the range of $G_0$ estimated from 8~$\mu$m emission (see Sec.~\ref{sec:8micron}). At the lowest ratios of $G_0/n \approx 40/900$, a relatively weak radiation field and high density, the \Htwo{} transition occurs at the lowest $A_V$, 0.45, and the CO-dark layer which the CRRLs arise from has the largest $\Delta A_V(\mathrm{dark})$, 0.68. At the highest ratios of $G_0/n \approx 160/20$, a relatively strong radiation field and low density, the \Htwo{} transition occurs deeper into the cloud, 1.73, and the $\Delta A_V(\mathrm{dark})$ is a bit smaller, 0.46.

Prior low-frequency CRRL investigations estimated
$A_V \approx 0.3 - 1$ \citep[e.g.,][]{Salas2018} using the CRRL derived physical properties together with PDR models. Values
closer to $A_V \approx 2-3$ under the more extreme conditions found in Cyg X, $G_0 \simeq 100$ and $P \approx 10^5 - 10^6$~K~\cmc{} are predicted by PDR modeling \citep{LePetit2006, LeBourlot2012, Bron2016}. These $A_V$ are roughly inline with the values presented in Table~\ref{tab:physprops}.

\begin{figure}[!t]
    \centering
    \includegraphics[width=0.8\linewidth]{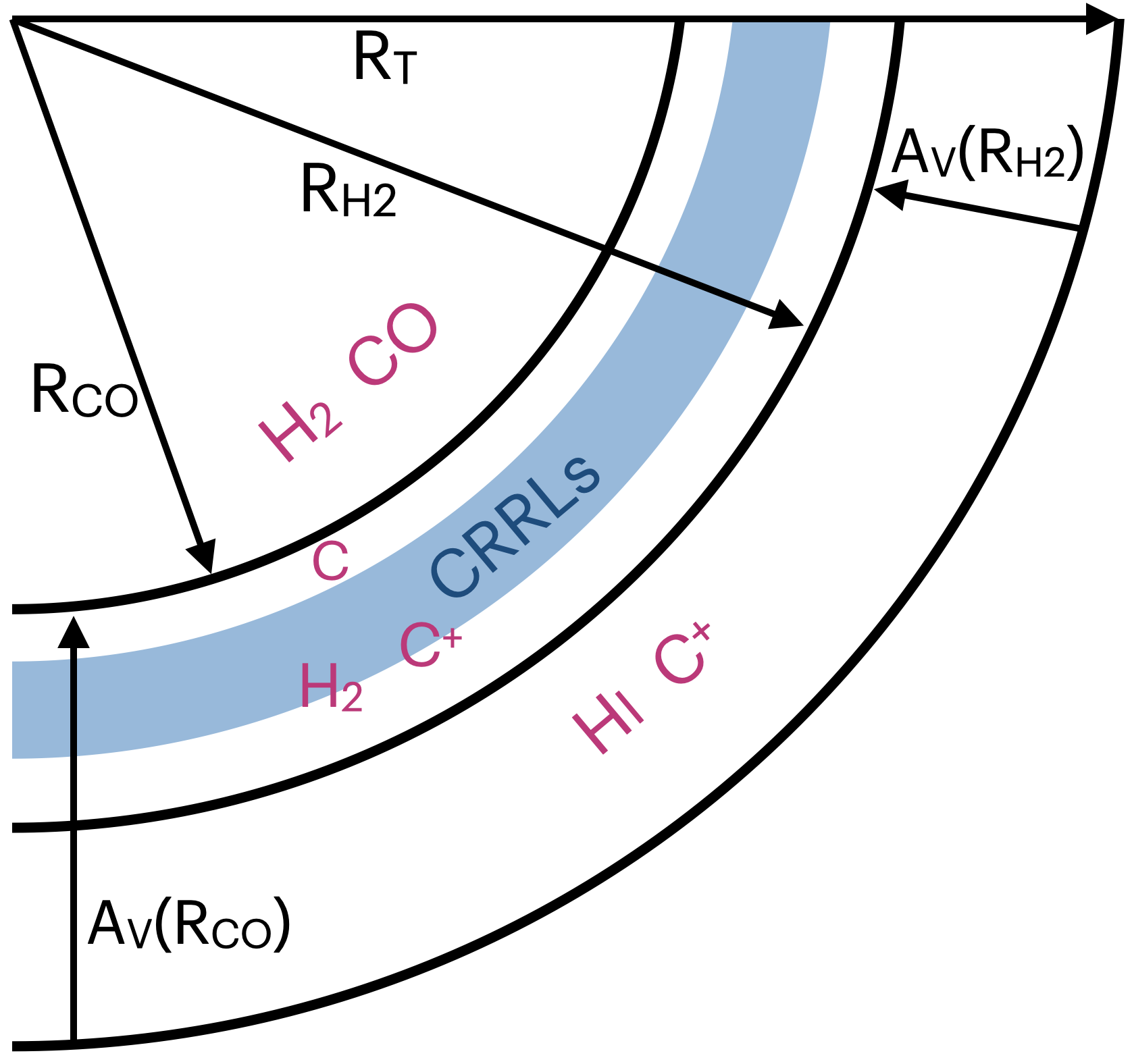}
    \caption{Illustration of a model cloud showing the parameters defined in the \citet{Wolfire2010} framework. $R_{\mathrm{CO}}$ is the radius of the CO core. $R_{\mathrm{H}_2}$ is the radius where $2 n_{\mathrm{H}_2} = n_{HI}$ with equal mass density in H atoms and \Htwo{} molecules. $R_{\mathrm{T}}$ is the total cloud radius. Within $R<R_{\mathrm{CO}}$, gas is mainly CO and \Htwo{}. Within the range $R_{\mathrm{CO}} < R < R_{\mathrm{H}_2}$, gas is mainly \Htwo{} whereas the gas-phase carbon is mainly C and C$^+$. The CRRLs arise from an C$^+$ layer that is mainly \Htwo{}.  Within $R_{\mathrm{H}_2} < R < R_{\mathrm{T}}$ gas is mainly \Hi{} and gas-phase carbon is mainly C$^+$. $A_V(R_{\mathrm{H}_2})$ is the magnitude of extinction measured from the outer radius to $R_{\mathrm{H}_2}$,and $A_V(R_{\mathrm{CO}})$ is the  magnitude of extinction measured from the outer radius to $R_{\mathrm{CO}}$. }
    \label{fig:wolfire_params}
\end{figure}

%%%%%%%%%%%%%%%%%%%%%%%%%%%%%%
\begin{deluxetable*}{CCCCCCCCCC}
    \tablecaption{Gas physical properties as estimated using the \citet{Wolfire2010} framework. \label{tab:physprops}}
    \tablewidth{0pt}
    \tablehead{
    \colhead{$\int I_{^{13}CO} dv$} %1
    & \colhead{$R_{\mathrm{CO}}$} %2
    & \colhead{$R_{\mathrm{H_2}}$} %3
    & \colhead{$\Omega_{\mathrm{CO}}$} %3
    & \colhead{$N_{\mathrm{H}_2}$} %4
    & \colhead{$n$} %5
    & \colhead{$G_0$} %6
    & \colhead{$A_V(R_{\mathrm{H}_2})$} %7
    & \colhead{$A_V(R_{\mathrm{CO}})$} %8
    & \colhead{$\Delta A_V(\mathrm{dark})$} \\ %9
    \colhead{[K~km~s$^{-1}$]}  %1
    & \colhead{[pc]} %2
    & \colhead{[pc]} %3
    & \colhead{} %3
    & \colhead{[cm$^{-2}$]} %4 
    & \colhead{[cm$^{-3}$]} %5
    & \colhead{} %6
    & \colhead{} %7
    & \colhead{} %8
    & \colhead{} %9
    }
    \startdata
    10  & 12 & 14.4 & 0.33 & 1.6 \times 10^{22} & 500 & 40  & 0.59 & 1.25 & 0.66 \\
        &    &      &      &                    &     & 160 & 0.93 & 1.53 & 0.60 \\
        & 10 & 12 & 0.23 & 2.3 \times 10^{22} & 900 & 40  & 0.45 & 1.13 & 0.68 \\
        &    &      &      &                    &     & 160 & 0.79 & 1.41 & 0.62 \\
    5   & 12 & 14.4 & 0.33 & 7.8 \times 10^{21} & 300 & 40  & 0.72 & 1.35 & 0.63 \\
        &    &      &      &                    &     & 160 & 1.06 & 1.64 & 0.58 \\
        & 10 & 12 & 0.23 & 1.1 \times 10^{22} & 500 & 40  & 0.59 & 1.25 & 0.66 \\
        &    &      &      &                    &     & 160 & 0.93 & 1.53 & 0.60 \\
    1   & 21 & 25   & 0.33 & 1.6 \times 10^{21} & 20  & 40  & 1.39 & 1.91 & 0.52 \\ 
        &    &      &      &                    &     & 160 & 1.73 & 2.19 & 0.46 \\
    \enddata
    \tablecomments{$\int I_{^{13}CO} dv$ is the velocity-integrated \treceCO{} line flux. $R_{\mathrm{CO}}$ is the radial extent of CO-traced emission. $R_\mathrm{H_2}$ is the radial extent of $H_2$ gas. $\Omega_{^{13}\mathrm{CO}}$ is the beam filling factor of the \treceCO{} emission. $N(\mathrm{H_2})$ is the \treceCO{}-traced $H_2$ column density. $n$ is the H nucleus density of the clumps. $G_0$ is the FUV radiation field in multiples of the average interstellar radiation field. $A_V(R_{\mathrm{H_2}})$ is the extinction measured from the outer surface of the clump to $R_\mathrm{H_2}$. $A_V(R_{\mathrm{CO}})$ is the extinction measured from the outer surface of the clump to $R_\mathrm{CO}$. $A_V(\mathrm{dark}) = A_V(R_{\mathrm{CO}}) - A_V(R_{\mathrm{H_2}})$.  }
\end{deluxetable*}

%%%%%%%%%%%%%%%%%%%%%%%%%%%%%%%%%%%%%%%%%%%%%%%%%%%%
\subsection{Pressure estimates} 
\label{ssec:region_props}

We estimate and compare pressure terms in order to understand what dominates the dynamics and evolution of the cool dark gas. In summary, we find that the turbulent pressure likely dominates on these scales. The turbulent pressure is given by $P_{\mathrm{turb}} = \rho V_{\mathrm{rms}}^2$ where $\rho$ is the mass density and the rms velocity dispersion, $V_{\mathrm{rms}}$ is related to the line-of-sight velocity dispersion as $V_{\mathrm{rms}} = \sqrt{3} \sigma_v$. For a nominal electron density of the \refn{} gas of $n_e \sim 0.06$~\cmc{} ($n_H \sim 400$~\cmc{}), the mass density is estimated as $\rho_{\mathrm{HI}} \approx 1.36 m_{\mathrm{H}} (n_e / A_c)$ and/or $\rho_{\mathrm{H_2}} \approx 2.36 m_{\mathrm{H}} (n_e / A_c)$. With $A_c =1.4 \times 10^{-4}$ \citep{Sofia2004} and $V_{\mathrm{rms}} = 7.8$~\kms{}, the turbulent pressure is $P_{\mathrm{turb}}/k_B \approx (\frac{n_e}{0.06~\mathrm{cm^{-3}}}) 4.2 - 7.2 \times 10^{6}$~\cmc{}~K.

In comparison, the thermal pressure of the CRRL gas is small,
where $P_{\mathrm{therm}} = k_B (n_e / A_c) T_e$. For gas with $n_e = 0.06$}~\cmc{} and $T_e = 50$~K, the thermal pressure is $P_{\mathrm{therm}} / k_B \approx (\frac{n_e}{0.06~\mathrm{cm^{-3}}}) 2.0 \times 10^{4}$~\cmc{}~K. In comparison, the ram pressure imparted by the \refn{} emitting gas is given by $P_{\mathrm{ram}} = \rho V^2$ where $\rho$ is the mass density and $V$ is the velocity. Assuming a nominal electron number density $n_e \sim 0.06$~\cmc{} and velocity of $V = 2.9$~\kms{}, the estimated ram pressure is $P_{\mathrm{ram}}/k_B \approx (\frac{n_e}{0.06~\mathrm{cm^{-3}}}) (6.0 - 10.2) \times 10^{5}$~\cmc{}~K.

\citet{Emig2022} investigated diffuse ionized gas structures throughout Cygnus X and estimated its thermal pressure to typically be $P_{\mathrm{ion}}/k_B \approx 6 \times 10^{5}$~\cmc{}~K, which was also in agreement with the X-ray studies $P_{X}/k_B \approx 6 \times 10^{5}$~\cmc{}~K.  Higher pressures are associated with compact \Hii{} regions such as DR21. The magnetic pressure is on the order of $P_{\mathrm{mag}}/k_B \approx 2 \times 10^{5}$~\cmc{}~K, assuming the magnetic field strength of $B=0.1$ mG in the ambient ISM near DR21 \citep{Ching2022}; higher turbulent pressures than magnetic pressures were also found with the CRRL observations towards Cas A \citep{Oonk2017}.

We estimate the eddy turnover time of a turbulent medium, $t_{\mathrm{turb}} = L / (V_{\mathrm{rms}})$, for the given size scales and line widths that we derive for the \refn{}-emitting gas. Taking $L = 21$~pc, equivalent to the beam size, and $V_{\mathrm{rms}} = 7.8$~\kms{} as above, the eddy turnover time is 2.6~Myr. In turbulent ISM models of molecular cloud formation, this sets an upper limit for the timescale of \Htwo{} formation, which does not depend on the nature of the turbulence present \citep[e.g.,][]{Micic2012}.  With an \Htwo{} formation timescale that is dependent upon density as $t_{H_2} \approx (10^3\,\mathrm{Myr}) / (n_{\mathrm{H}}\,\mathrm{[cm^{-3}]})$ \citep{Hollenbach1971b}, 2.6~Myr estimate implies the density in the gas forming \Htwo{} is $n_{\mathrm{H}} \sim 380$~\cmc{}, or $n_e \sim 0.05$~\cmc{}. The density we have roughly estimated by way of the line widths is quite comparable to the density estimated in Section~\ref{ssec:discuss_spatialsep} from the typically spatial separation of \refn{} and \treceCO{}.

Could the velocity difference of \treceCO{} with the \refn{} components (Section~\ref{sssec:vel_offset}) represent free-fall velocities from gravitational collapsing material throughout the region, as seen towards DR21 and W75N (see Section~\ref{ssec:discuss_dr21})? Combining the 2.9~\kms{} velocity flow with the average spatial separation of 12~pc (Section~\ref{ssec:spatial_sep}), the timescale for the convergence of \refn{} material onto the molecular cloud is $t \sim 4$~Myr. This sits above the turbulent timescale of 2.6 Myr. Turbulence may appreciably act on the gas and form \Htwo{} on spatial and time scales shorter than the build-up from the flow of material onto the cloud; the same can be said for the disruption of \Htwo{} and possible flow of material away from the cloud. However, given that the nominal \refn{} line width could have a large systematic uncertainty, confirmation via a better characterization of the line profiles is worthwhile.

%%%%%%%%%%%%%%%%%%%%%%%%%%%%%%%%%%%%%%%%

%%%%%%%%%%%%%%%%%%%%%%%%%%%%%%%%%%%%%%%%%%%%%%%%%%%%%%
\subsection{Forming Molecular Gas in the DR21 region: comparison with \Cii{}} 
\label{ssec:discuss_dr21}

The DR21 and W75N regions in Cygnus X are iconic massive star-forming regions which have been extensively studied in the literature \citep{Reipurth2008}. The dense filaments of molecular gas associated with each of these regions  are undergoing gravitational collapse \citep{Schneider2010, Li2023, Zeng2023}. The DR21 and W75N clouds are colliding head-on \citep{Dickel1978, Dobashi2019} and/or are forming from the interaction of composite \Hi{} and \Htwo{} clouds \citep{Schneider2023, Bonne2023}.  The slightly closer W75N cloud, at $1.3 \pm 0.1$~kpc \citep{Rygl2012}, is moving away from the observer, with a red-shifted systemic velocity of $\sim$9~\kms{}. The slightly more distant DR21 cloud, at $1.5 \pm 0.1$~kpc \citep{Rygl2012}, is moving towards the observer with a blue-shifted systemic velocity of $-$3~\kms{} (see Figure~\ref{fig:DR21}). A molecular component centered at 3.5~\kms{} could be emission bridging the clouds as a result of the collision \citep[e.g.,][]{Haworth2015, Dobashi2019} and/or related to a foreground ($d \sim 600-800$~pc) cloud, the Cygnus Rift \citep[e.g.,][]{Gottschalk2012}.

%%%%%%%%%%%%%%
\begin{figure}
    \centering
    \includegraphics[width=0.47\textwidth]{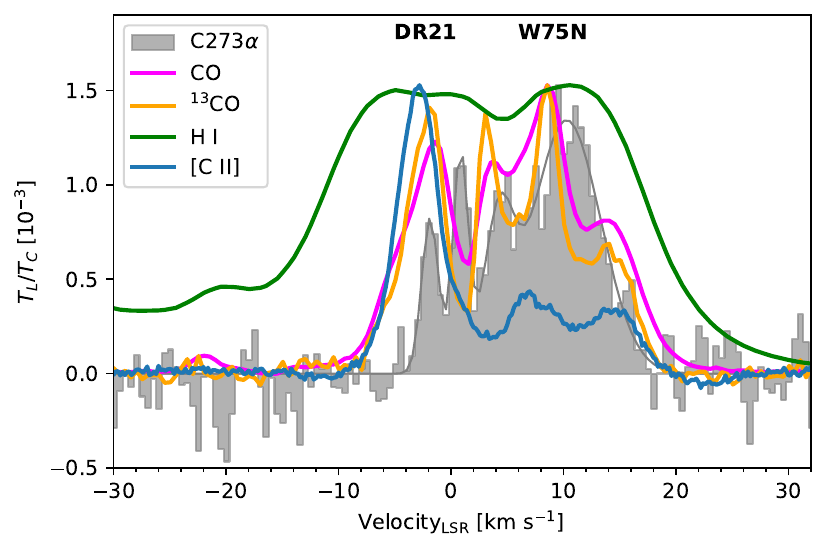}
    \caption{Spectra encompassing well-known DR 21 ($\sim -3$~\kms{}) and W75N ($\sim +9$~\kms{}) regions in Cyg X North, the region in our footprint for which spectrally-resolved \Cii{} 158 $\mu$m data have been taken. Spectra of \Hi{}, \CO{}, \treceCO{}, and \Cii{} have been normalized to the peak of \refn{}.}
    \label{fig:DR21}
\end{figure}
%%%%%%%%%%%%

A footprint covering DR21 and W75N is the only region in our \refn{} map for which \Cii{} 158~$\mu$m emission has so far been observed at high spectral resolution, thanks to the SOFIA legacy program FEEDBACK \citep{Schneider2020}.Using the SOFIA FEEDBACK data, \cite{Schneider2023} investigated the low excitation \Cii{} emission from W75N and a high velocity component ($v_{\mathrm{LSR}} = 4-20$~\kms{}), and \cite{Bonne2023} investigated the low excitation \Cii{} associated with DR21 at $v_{\mathrm{LSR}} \lesssim 0$~\kms{}.

[CII] 158~$\mu$m emission is an interesting comparison point because [CII] and low-frequency CRRLs can indeed arise from the same gas volume \citep[e.g.,][]{Salas2019, Salgado2017b}. However, in addition to cool dark gas, [CII] also arises from other (denser and warmer) gas phases \citep[e.g.,][]{Pineda2013}. Identifying [CII] from cool dark gas requires a decomposition of emission which is low in intensity compared to [CII] from dense PDRs \citep[e.g.,][]{Bonne2023}. Although low-frequency CRRLs are comparatively weak in strength, they do arise exclusively from cool dark gas and provide an uncontaminated view of this gas phase.

The SOFIA FEEDBACK data covers a 0.26 deg$^{2}$ footprint centered about $(\ell, b) = (81.605^{\circ}, +0.559^{\circ})$. In Figure~\ref{fig:DR21}, we plot the \Cii{} spectrum averaged over the FEEDBACK footprint. We also overlay the \refn{}, \Hi{} 21 cm, \CO{}, and \treceCO{} emission that has been extracted from a single pixel of 48\amin{} resolution data. Our beam size equates to an area of 0.45 deg$^{2}$ and is a bit larger than the \Cii{} footprint. 

Four Gaussian components fit to the \refn{} spectrum result in the lowest Bayesian and Akaike information criteria, in comparison to one, two, three, five, or six component fits. The properties of the best-fit profiles are given in Table~\ref{tab:dr21_lines} and plotted in Figure~\ref{fig:DR21}. These components are generally separated in velocity from CO components, by 2--3~\kms{}. They are all on the narrow end of the line-width distribution determined for our full survey data (Section~\ref{sec:results}), with the higher S/N in this location seemingly helps to discern narrower profiles. 

%%%%%%%%%%%%%%%%%%%%%%%%%%%%%%
\begin{deluxetable}{C C C C C}
    \tablecaption{\refn{} Line Profiles of DR21/W75N (see Figure~\ref{fig:DR21})
    \label{tab:dr21_lines}}
    \tablewidth{0pt}
    \tablehead{
    \colhead{} 
    & \colhead{V$_{cen}$} %2
    & \colhead{Peak} %3
    & \colhead{FWHM}
    & \colhead{} \\ %5
    \colhead{}  & \colhead{[\kms]} & \colhead{[$T_L/T_C~10^{-3}$]} & \colhead{[\kms]} & \colhead{} 
    }
    \startdata
    & -1.86 \pm 0.16 & 0.83 \pm 0.14 & 1.74 \pm 0.38 & \\ 
    & 0.84 \pm 0.12 & 1.15 \pm 0.15 & 1.62 \pm 0.54 & \\ 
    & 4.24 \pm 0.27 & 0.76 \pm 0.13 & 3.06 \pm 0.80 & \\ 
    & 10.27 \pm 0.27 & 1.34 \pm 0.07 & 7.11 \pm 0.68 & \\ 
    \enddata
    \tablecomments{Best-fit Gaussian properties: ``V$_{\mathrm{cen}}$'' is the central velocity, and ``Peak'' is the peak amplitude, and ``$\sigma_{v}$'' is the Gaussian width. }
\end{deluxetable}

The \refn{} $-1.9$~\kms{} component coincides with \Cii{} emission from the cold \Hi{} and molecular subfilaments of DR21, falling predominately over -3 to -1~\kms{} as characterized at high resolution (14\asec, 0.1~pc) by \citet{Bonne2023}. In Figure~\ref{fig:DR21}, this low excitation \Cii{}-emitting gas is confused by the bright \Cii{} associated with the DR21 high density PDR ($v_{LSR} \sim -3$~\kms{}). At 0.1 pc resolution, the low-excitation \Cii{} line widths were found to be 4.0--5.0~\kms{}; the \Cii{} appears as a thin sheet of approximate density $n \approx 5000$~\cmc{}, embedding molecular subfilaments that are about 0.3 pc in size \citep{Hennemann2012} and have molecular line widths of $<1.3$~\kms{}. Even in the 48\amin{} GBT beam, the \refn{} line-widths are smaller than those of \Cii{}. The \refn{} gas is therefore likely colder than the \Cii{} gas. The warmer gas which emits at 158 $\mu$m has larger turbulent and bulk motions, while the cooler gas emitting in the CRRL is considerably more quiescent. Either a $\sim$single coherent \refn{} component dominates even on large scales, or if Doppler broadening of multiple components is present at this scale, then the line widths of individual components are intrinsically smaller. In either case, \refn{} reasonably has a higher molecular gas fraction than \Cii{}. Low-frequency CRRL emission towards Cas A was also found to have a high molecular gas fraction  \citep{Salas2018}. With $\ell_{CO-dark} \approx 0.2 \cdot \ell_{CO}$ \citep{Wolfire2010}, the pathlength of (a collection of) \refn{} layers could approach $\sim0.1$~pc.

\cite{Bonne2023} performed a detailed comparison of the gravitational potential, magnetic field, and turbulent support in DR21, and determined that the gravitational energy dominates. The velocity offset of \Cii{}, 1--2~\kms{}, is attributed to gravitational collapse. Infalling molecular gas in the region has smaller velocity differences of $\sim$0.6~\kms{} \citep{Schneider2010}. The velocity offset of \refn{} melds well into this picture.

Furthermore, we also reason that \refn{} emitting gas is inflowing onto the DR21 cloud. \refn{} is only observable when illuminated by background continuum emission. DR21 is more distant within the Cygnus X complex than W75N, and  so the \refn{} emission we observe likely falls in regions of the cloud that are on the near side of the main DR21 cloud. 
Therefore, since the \refn{} emission is at comparable and only red-shifted velocities with respect to the DR21 cloud's systemic velocity, the bulk motion of \refn{} would inflow towards the cloud. This indicates that \refn{} is tracing cold-dark material accreting onto the cloud.

Like DR21, the W75N cloud, at $8.6 \pm 0.25$~\kms{}, is known to be experiencing gravitational collapse which dominates over magnetic support \citep{Zeng2023}. The brightest \refn{} component at 10.3~\kms{} seems most likely to be associated with W75N. In Figure~\ref{fig:DR21}, a velocity offset between the \refn{} and \CO{} of about 1.7~\kms{} is present, red-shifted. Molecular gas within and connected to the W75N area is known to span a large velocity range, with brightest emission over a gradient of 8 to 11~\kms{} \citep{Dickel1978, Schneider2006}. \cite{Schneider2023} also find low excitation \Cii{} gas that spans 4--12~\kms{}.

%%%%%%%%%%%%%%%%%%%%%%%%%%%%%%%%%%%%%%%%%%%%%%%%%%%%%%
\subsection{Size Scales of Cool Dark Gas from \refn{}} 
\label{ssec:discuss_size}

Cool dark gas in our Galaxy is distributed anisotropically \citep{Heiles2003b, Heiles2005, Grenier2005}. Cool \Hi{} in emission \citep{Clark2014} and absorption \citep{McClure-Griffiths2006} reveal ubiquitous filamentarity. These narrow ($\sim$0.1 pc) \citep{Clark2014, Kalberla2016} density structures \citep{Clark2019a} have high aspect ratios \citep[$\gtrsim 100$;][]{Kalberla2023}. Their orientation appears to be parallel to magnetic fields at column densities of $N(H) < 5 \times 10^{21}$~cm$^{-2}$ and perpendicular at higher column densities \citep{McClure-Griffiths2006, Clark2015, Planck2016XXXV, Kalberla2020}. While these structures are characterized by the presence of \Hi{}, they may also contain predominantly molecular gas \citep{Kalberla2020}. \citet{Strasser2007} estimate that the mean distance between cold absorbing clouds is 90--220 pc.

Low-frequency CRRLs have been directly connected to HISA from cool dark filaments in the Sun's local bubble \citep{Roshi2011}. \citet{Roshi2011} estimated line-of-sight CRRL path lengths in the range 0.03--3.5~pc in the Riegel-Crutcher Cloud at a spatial resolution of $4.4 \times 1.3$ pc$^2$ (2\deg{}~$\times$~0.6\deg{}). These path lengths agree with those of the CNM filaments in swept up shells. Towards another line-of-sight, the supernova remnant Cassiopeia A, CRRLs associate with a large molecular cloud in the Perseus spiral arm. CRRL analyses of two bright gas components with LOFAR and WSRT have determined path lengths of the CRRL-emitting gas with small uncertainty. \citet{Oonk2017} determined a path length of $35.3 \pm 1.2$~pc for the gas across the 6\amin{} (5.5~pc) extent of Cas A. \citet{Salas2018} resolved this region at 70\asec{} (1.0~pc) to directly show the CRRLs tracing the surface of the CO-cloud, with projected spatial separations between \CO{} and CRRL emission of 1--2~pc. Their line-of-sight integrated path lengths varied between 27--182~pc for a single velocity component. \citet{Chowdhury2019} used GMRT 430 MHz observations at 18\asec{} (0.3 pc) resolution which show point-like condensations and linear-like bright emission that have unresolved widths ($<0.3$~pc) and a linear extent of $>3$~pc. The single velocity component indicates a coherent gas structure, but the long path lengths indicate sheet-like CNM structure, which contrasts with filament structure as determined from \Hi{} and dust emission analyses.

Cygnus X contrasts starkly with the local \Hi{} cloud structures and the random-chance sight-lines probed by the observations discusses above, as well as with the diffuse ISM towards Cas A observed in low-frequency CRRLs. The gas pressures in Cyg X are higher by at least an order of magnitude \citep[][see also Section~\ref{ssec:region_props}]{Emig2022}. More than 170 OB stars contribute to a high radiation field and are actively churning up their molecular cloud environment, blowing bubbles, illuminating, photo-evaporating, photo-ionizing molecular cloud edges, and pushing gas around onto and away from clouds. The \refn{} observations presented in this article provide a window into cold gas structures in the presence of these feedback activities.

The \refn{} morphology indeed shows elongation (filamentarity) of cool dark gas, similar to previous works. However, as best as can be discerned with the spatial resolution available, this elongation spans larger extents than previously reported, more than 100 pc and contains numerous spatio-spectral coherent structures. The mean separation inferred between these ``clouds'' is thus considerably smaller while their coherent extent is larger than previous works \citep{Strasser2007, Bellomi2020}. It could be that some of the structures are exposed cloud cores or long, thin illuminated cloud edges \citep[e.g.,][see also Sections~\ref{ssec:discuss_DR4}~\&~\ref{ssec:ngc6913}]{Emig2022}.

%%%%%%%%%%%%%%%%%%%%%%%%%%%%%%%%%%%%%%%%%%%%%%%%%%%%%%%%%%%%%%%%%
\subsection{Supernova Remnant G78.2+2.1 and the NGC 6910 cluster} 
\label{ssec:discuss_DR4}

%%%%%%%%%%%%%%
\begin{figure}
    \centering
    \includegraphics[width=0.7\linewidth]{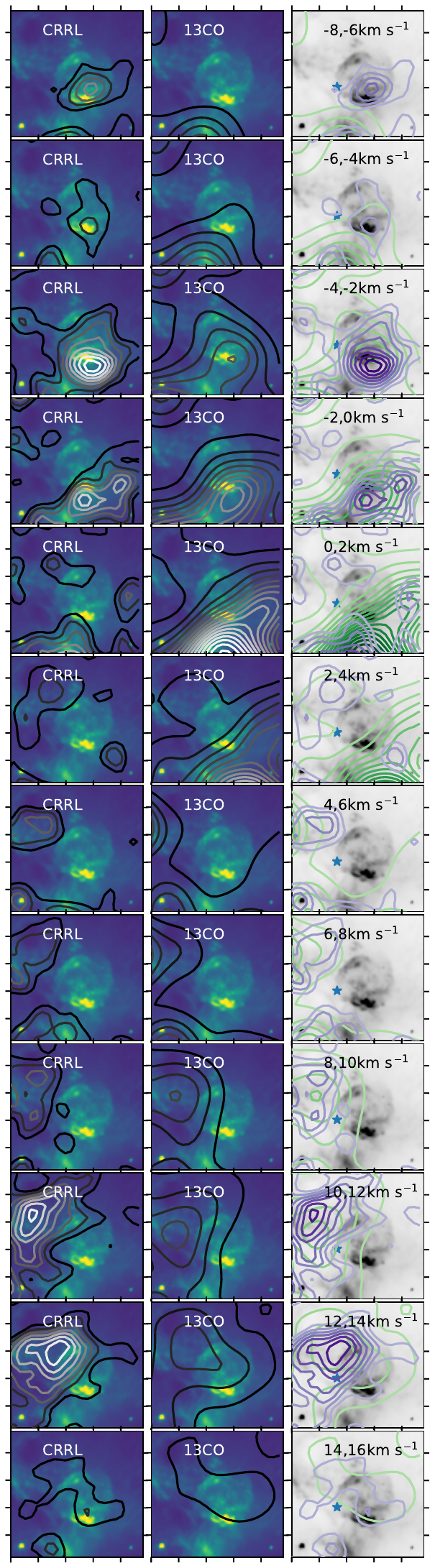}
    \caption{Emission in the region of the SNR $\gamma$-Cygni and cluster NGC~6910 (blue star; \textit{right} column). 1.4~GHz continuum emission is the background image. Maps integrated over 2~\kms{} show \refn{} contours (\textit{left} column), \treceCO{} contours (\textit{middle} column), and with both \refn{} (purple) and \treceCO{}  (green) contours overlaid in the \textit{right} column.}
    \label{fig:SNR_chanmaps}
\end{figure}

The shell-type supernova remnant (SNR) G78.2+2.1 has a distance of about 1.8~kpc \citep{Higgs1977b}. It has been associated with the $\gamma$-Cygni nebula \citep{Higgs1977b}, the radio continuum sources DR3 and DR4 \citep{Downes1966}, and with the pulsar PSR J2021+4026 \citep{Trepl2010}. \Hi{} absorption and emission features have been investigated for interaction and influence with the SNR \citep{Landecker1980, Braun1986, Gosachinskij2001, Ladouceur2008, Leahy2013}. \Hi{} absorption against the SNR continuum is detected at least at the LSR velocities of -8~\kms{} to +20~\kms{}, also the approximate range covered by the molecular gas in the region. It is clear that the SNR is at a distance further than the material associated with these velocities, but discerning which components possibly associate with the SNR interaction is challenging, leading to multiple scenarios proposed \citep[e.g.,][]{Ladouceur2008}.

\refn{} is observed from about -8 to 16~\kms{} in a 2.5\deg{}~$\times$~2.5\deg{} region in this direction (Figure~\ref{fig:SNR_chanmaps}). \cite{Roshi2022} observed DR4 at 321 and 800 MHz with the GBT, and characterized a single velocity component in this direction ($v_\mathrm{LSR} = -2.7 \pm 0.3$~\kms{}), albeit with 4~\kms{} channel resolution. They noted the CRRL central velocity coincides with the velocity of an \Hi{} self-absorption component in CGPS data. 

With the larger area mapped here and at high spectral resolution, we find additional components and image their kinematics. The kinematics of the blue-shifted \refn{} emission, see Figure~\ref{fig:SNR_chanmaps}, are consistent with an expanding shell moving at (at least) $\sim$8--10~\kms{}, with observed central velocities of -9 \kms{} up to, possibly, +1~\kms{}. Comparable scenarios, including some with larger expansion velocities, have been proposed \citep{Landecker1980, Ladouceur2008, Leahy2013}.
\refn{} emission appears as unique spatio-spectral peaks, at -7~\kms{} and -3~\kms{}. These may be two condensations of molecular gas. The spectral profiles show the asymmetry expected of an expanding shell of material, with lower level emission at the most extreme velocities peaking closest to the projected center of the supernova remnant shell and with the brightest emission closest to the shell edges in projection.  The \refn{} component at -8 to -6 ~\kms{} is coincident with the most blue-shifted components of \Hi{} absorption at $\sim$-8~\kms{} \citep{Leahy2013}.

The morphology of the emission at -1~\kms{} is elongated at the Southern rim of the supernova remnant. And in the 1~\kms{} channel, three regions surrounding the shell also appear in \refn{} emission. However, emission at -1~\kms{} to 3~\kms{} shows morphologies possibly related to the massive star cluster NGC~6913, as described in Section~\ref{ssec:ngc6913}. In any case, it is interesting to note there are \refn{} emission peaks which surround \treceCO{} emission peaks in the  -1~\kms{} channel.

The red-shifted \refn{} emission must be in front of the supernova remnant, otherwise the \refn{} strength (as a line-to-continuum ratio) would be significantly diluted by the strong continuum of the supernova remnant. Therefore the redshifted \refn{} components should not be related to the receding side of an expanding shell, at odds with the scenarios that suggest this association \citep{Ladouceur2008}. The redshifted \refn{} emission extends to about 15~\kms{}, velocities that \cite{Landecker1980} suggested harbor a cold \Hi{} screen. The \refn{} data suggests that the cloud dynamics in this portion of the Cygnus X region may be projected onto but not directly related to an interaction with the SNR.

\refn{} and \treceCO{} emission start to appear in the NE of Figure~\ref{fig:SNR_chanmaps} from about 5~\kms{} and up. In channels 7-9~\kms{}, \refn{} appears as two local peaks again surrounding a \treceCO{} peak. In Channel 11~\kms{} \refn{} forms an arc like structure, and maintains brightness in the 13~\kms{} channel. The massive star-cluster NGC 6910, at a distance of  $1.7 \pm 0.1$~kpc \citep{Cantat-Gaudin2020, Quintana2022} may have some influence. It has a stellar mass of $10^3$~\Msun{} \citep{LeDuigou2002} and age of $6 \pm 2$~Myrs \citep{Kolaczkowski2004}. The arc in the 13~\kms{} channel may very well be related to photoionizing and/or stellar-wind feedback from the $\sim$30 OB stars making up NGC 6910 \citep{LeDuigou2002}.

%%%%%%%%%%%%%%%%%%%%%%%%%%%%%%%%%%%%%%%%%%%%%%
\subsection{The massive star-cluster NGC 6913} 
\label{ssec:ngc6913}

The PDR that extends in Figure~\ref{fig:cyg_intro} along DR4 and DR5 down to DR13 may be the PDR rim of a bubble powered by the massive star-cluster NGC 6913 \citep{Schneider2007}. NGC 6913 (M29) hosts about 20 OB stars \citep{LeDuigou2002}. \refn{} emission extending at negative velocities could relate to the rim of that stellar bubble. We show a \refn{} emission integrated over -4 to 2~\kms{} in Figure~\ref{fig:ngc6913}. Compressed, cooling gas behind the PDR rim could be rich in cold \Hi{} and/or CO-dark \Htwo{}.

%%%%%%%%%%%%%%
\begin{figure}
    \centering
    \includegraphics[width=0.2\textwidth]{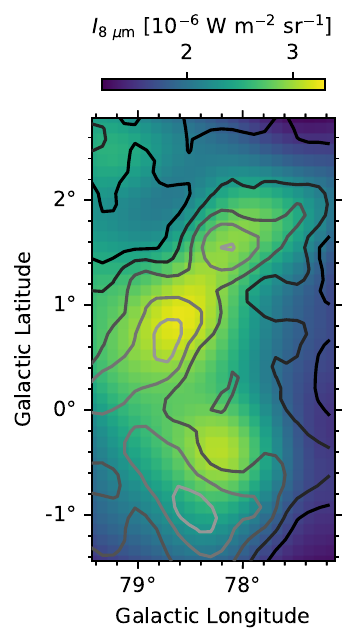}
    \caption{PAH emission at 8.3~$\mu$m with \refn{} integrated from -4 to 2 \kms{} overlaid, showing what a bubble powered by the cluster NGC 6913.}
    \label{fig:ngc6913}
\end{figure}

%%%%%%%%%%%%%%%%%
\section{Conclusions}
\label{sec:conclude}

We surveyed carbon radio recombination lines (CRRLs) at 292--394 MHz with GBT over a $5.5^{\circ}\,\times\,5.5^{\circ}$ area in the Cygnus X ($d \sim 1.5$~kpc) star-forming region. The low-frequency CRRL emission is predominantly stimulated and originates in cold ($T \lesssim 100$~K) gas where carbon is singly ionized.   We use these observations to investigate the \Hi{}-to-\Htwo{} transition. This article presents the first large-scale mapping \citep[$>$0.014\,deg$^2$; e.g.,][]{Salas2018} of low-frequency CRRLs from cool dark gas. 

By stacking up to 28 CRRLs, we created a line-synthesized data cube with an effective transition of \refn{} at an effective frequency of \refnu{} and spatial resolution of 48\amin{} (21 pc). We characterized the spatial distributions and spectral line profiles of \refn{} and compared the properties with those of \treceCO{} emission at matched resolutions.  A summary of our results are as follows:
\begin{enumerate}

\item We detect \refn{} emission in the Moment 0 map with a significance greater than 3$\sigma$ over 24\,deg$^2$ (75\%) of the mapped region (Figure~\ref{fig:cyg_intro}). The \refn{} emission spans LSRK velocities of $-13$~\kms{} to 17~\kms{} (Figure~\ref{fig:spec_spatialavg}), similar to \treceCO{}.

\item The morphology of the \refn{} is complex and varied. Arcs, linear-like ridges, and bright point-like condensations are visible throughout the region. Emission in a velocity channel (Figure~\ref{fig:chan_CRRL}) is typically extended and possibly sheet-like, and a given location typically has multiple velocity components present. Size scales of emission range from 10 pc (24\amin{}) to more than 100 pc. These findings contrast with previous reports of cool dark gas size scales. Some \refn{} bubbles and arcs may be related to the interaction of the SNR $\gamma$-Cygni with the surrounding ISM and stellar feedback bubbles surrounding the massive clusters NGC 6910 and NGC 6913.

\item To first order, locations with \refn{} emission are generally bright in \treceCO{}. The tracers can have similar morphologies (Figure~\ref{fig:chanmaps_CO}), with \refn{} sitting at the edges of peaked \CO{} emission. In other instances, \refn{} emission appears where \treceCO{} is relatively faint. At matched resolution, the \refn{} shows more structure, possibly indicating that it is not as wide-spread or volume-filling on all scales of \treceCO{} emission.

\item The \refn{} spectral profiles are well fit by Gaussians, indicating they are not collision or radiation broadened. They are likely to be turbulently broadened. The median FWHM$_{\mathrm{C}273\alpha }$ is $10.6$~\kms{}, but the line widths range from 2 -- 20~\kms{}. In comparison, the median line width fitted for \treceCO{} is FWHM$_{\mathrm{^{13}CO}} = 2.5$~\kms{}. The characteristic \refn{} line width implies Mach numbers of $\mathcal{M}_{\mathrm{C}273\alpha } \approx 10 - 30$ for $T \approx 20-100$~K which is significantly higher than found towards diffuse ISM sight lines. 
The \refn{} line widths imply this gas is dynamically more active than the \treceCO{} gas.

\item Velocity offsets between \refn{} and \treceCO{} are apparent throughout the region.  We compared the central velocity of each \refn{} fitted component with the \CO{} component closest in velocity (within the same spatial aperture). The velocity difference, $V_{\mathrm{C}273\alpha} - V_{\mathrm{^{12}CO}}$, has a standard deviation of 2.9~\kms{}. In the DR21/W75N region, the orientation and kinematics require \refn{} to be infalling towards the \treceCO{}-traced gas and thus likely in the process of forming molecular gas. 

\item We find a correlation with the \refn{} emission and 8~$\mu$m intensity, with a best-fit power-law slope of $1.3 \pm 0.2$. We estimate FUV radiation fields from the 8~$\mu m$ intensity, $G_0 \approx 40 - 165$. We interpret the relation as the dependence of cool dark gas emission on the FUV radiation field. The comparison of the \refn{} and \treceCO{} velocity-integrated emission shows perhaps mild correlation, with a normalization that possibly reflects local cloud conditions.

\item We characterize the angular separation between peaks of \refn{} emission and the closest peak of \treceCO{} emission in channel maps. The angular separation reveals a characteristic separation of 12~pc and a tail out to 30~pc. We use the \cite{Wolfire2010} framework to estimate that the CRRLs may arise from gas with densities of $n_H \approx 20 - 900$~\cmc{} and from cloud extinction layers of $A_V \approx 0.45 - 2.19$.

\end{enumerate}

The \refn{} emission likely arises from C$^+$/\Htwo{} gas, commonly referred to as CO-dark molecular gas. On these scales, the evolution of the \refn{} gas seems to be dominated by turbulent pressure, with a characteristic timescale to form \Htwo{} of about 2.6 Myr. However, higher S/N observations that more robustly characterize the line profiles (i.e., line widths) are needed to confirm this. Likewise, these data revealed a correlation between the low-frequency CRRLs and a proxy of the FUV radiation field for the first time. Extending these studies to cover a larger intensity regime and provide a longer leaver arm to assess the trends and distribution is needed. The GBT Diffuse Ionized Gas Survey at Low Frequencies (GDIGS-Low\footnote{\href{https://greenbankobservatory.org/science/gbt-surveys/gdigs-low/}{https://greenbankobservatory.org/science/gbt-surveys/gdigs-low/}}; PI: P.~Salas) is ideally suited to do this. GDIGS-Low is mapping the Inner Galaxy with RRLs at the 340 and 800 MHz windows. GDIGS-Low is an extension to the GDIGS survey of RRLs at 5.8 GHz \citep{Anderson2021}. Large systematic studies with GDIGS-Low and forthcoming observations with next-generation low-frequency telescopes will bring profound insights into cool dark gas using ionized carbon lines at low-frequencies. These observations highlight the GBT and the 340 MHz window \citep[e.g., see also,][]{Anantharamaiah1985a, Roshi1997, Roshi2002} as an excellent probe for CRRL studies.

%%%%%%%%%%%%%%%%%%%%%%%%%%%%%%%%%%%%%%%%

\acknowledgments

We thank the referee for their time and efforts with a helpful and insightful review.
KLE and PS dedicate this article to Violeta Emig Salas.
The Green Bank Observatory is a facility of the National Science Foundation operated under cooperative agreement by Associated Universities, Inc.
The National Radio Astronomy Observatory is a facility of the National Science Foundation operated under cooperative agreement by Associated Universities, Inc.  
This research made use of astrodendro, a Python package to compute dendrograms of Astronomical data (\href{http://www.dendrograms.org/}{http://www.dendrograms.org/}).
NS acknowledges support by the Federal Ministry of Economics and Energy (BMWI) via DLR, Projekt Number 50 OR 2217 (FEEDBACK-plus) and the Collaborative Research Center 1601 (sub-project B2) funded by the DFG, German Research Foundation) – 500700252.
GJW gratefully acknowledges the receipt of an Emeritus Fellowship from The Leverhulme Trust.
RCL acknowledges partial support for this work provided by a National Science Foundation (NSF) Astronomy and Astrophysics Postdoctoral Fellowship under award AST-2102625.
This project was supported by the National Aeronautics \& Space Administration through the University of Central Florida’s (UCF) Florida Space Institute and UCF’s NASA Florida Space Grant Consortium and Space Florifda, Grant Number 80NSSC20M0093.

\facilities{GBT, Effelsberg, DRAO, FCRAO, MSX, PMO, SOFIA}


\software{
Astrodendro \citep{Rosolowsky2008},
Astropy \citep{Astropy2018, Astropy2022},
CARTA \citep{Comrie2021},
CRRLpy \citep{Salas2016},
GaussPy+ \citep{Riener2019},
Matplotlib \citep{Hunter2007},
and NumPy \citep{Harris2020}
}

%%%%%%%%%%%%%%%%%%%%%%%%%%%%%%%%%%%%%%%%
\appendix


\section{Noise Estimation}
\label{ap:sec:noise}

We produce a three-dimensional estimate of the rms noise. Our methodology closely follows that of \citet[][see their Section 7.2]{Leroy2021}. We first construct a noise map that captures spatial variations of the noise, $R(x,y)$. Then, we measure a normalized noise spectrum that captures relative spectral variations, $s(v)$. The noise in the data cube is then 
\begin{equation}
\sigma(x,y,v) = R(x,y)s(v).
\end{equation}

We determine $R(x, y)$ and $s(v)$ empirically, determining the values from the data themselves using an iterative procedure. First, we compute spatial variations. We use the line-free channels 800--1200 (see Fig.~\ref{ap:fig:noise}) and calculate the standard deviation at each pixel, yielding an estimate of $R(x, y)$. Next, we compute spectral variations. We take all spatial pixels of a given channel. Then we exclude positive data that have high significance with respect to the noise level. These are likely to be associated with real emission and not noise. We do not exclude high-significance negative data. We then calculate the standard deviation for each channel while weighting each pixel with its spatial response $R(x,y)$. We normalize the noise standard deviation of each channel by the mean value. We then smooth these estimates with a third-order Savitsky–Golay filter to estimate $s(v)$.

The calculation of $R(x, y)$ is then repeated, but this time weighting each of the line-free channels by the estimate of $s(v)$. Then we repeat the calculation for $s(v)$. Finally, we iterate once more on the procedure to ensure convergence, and multiply the final $R(x, y)$ and $s(v)$ responses to generate an estimate of $\sigma (x,y,v)$. The iterative process drives the signal-to-noise distribution, $\frac{T_{\mathrm{L}}}{T_{\mathrm{C}}}(x, y, v)/ \sigma(x, y, v)$, in the signal-free regions to a normal distribution centered on zero with a standard deviation of 1. 

Figure~\ref{ap:fig:noise} shows the empirically derived noise and the signal-to-noise distribution. The weighted mean of the noise is $\sigma_{T_L / T_C} = 1.9 \times 10^{-4}$ in 0.5~\kms{} channels.

\begin{figure}[H]
    \centering
    \includegraphics[]{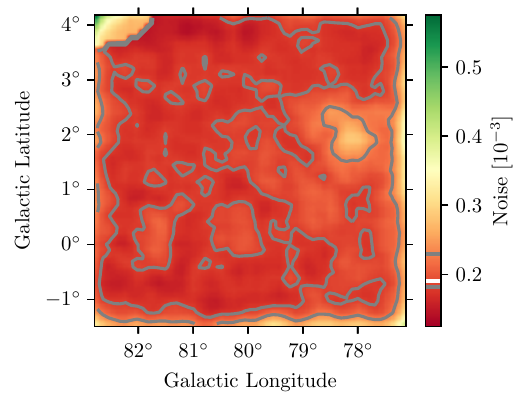}\\
    \includegraphics[width=0.48\textwidth]{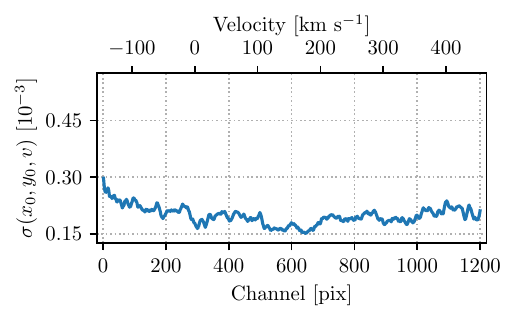}
    \includegraphics[width=0.48\textwidth]{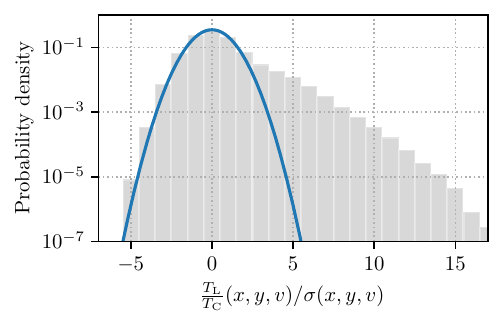}\\
    \caption{The top and middle panels show the noise along the spatial (top) and spectral (middle) coordinates. Gray contours in the top panel show the 50th and 90th percentiles of the noise values and the mean, $1.91 \times 10^{-4}$, is indicated by a white line in the colorbar. The noise profile along the spectral axis (middle panel) is extracted from the center of the map. The bottom panel shows the probability density function of the signal-to-noise within the cube. The blue parabola shows the PDF of a normal distribution with mean of zero and variance of one. A normal distribution is an excellent description of the signal-to-noise values, except for the strong positive tail of values arising from signal in the cube.}
    \label{ap:fig:noise}
\end{figure}

%%%%%%%%%%%%%%%%%%%
\section{Map Comparisons at Matched Spatial Resolutions}
\label{ap:sec:map_matchedspatial}

We show maps of various tracers compared with velocity-integrated \refn{} emission in Figure~\ref{ap:fig:overlay_matched_res}.

\begin{figure}[H]
    \centering
    \includegraphics[width=\linewidth]{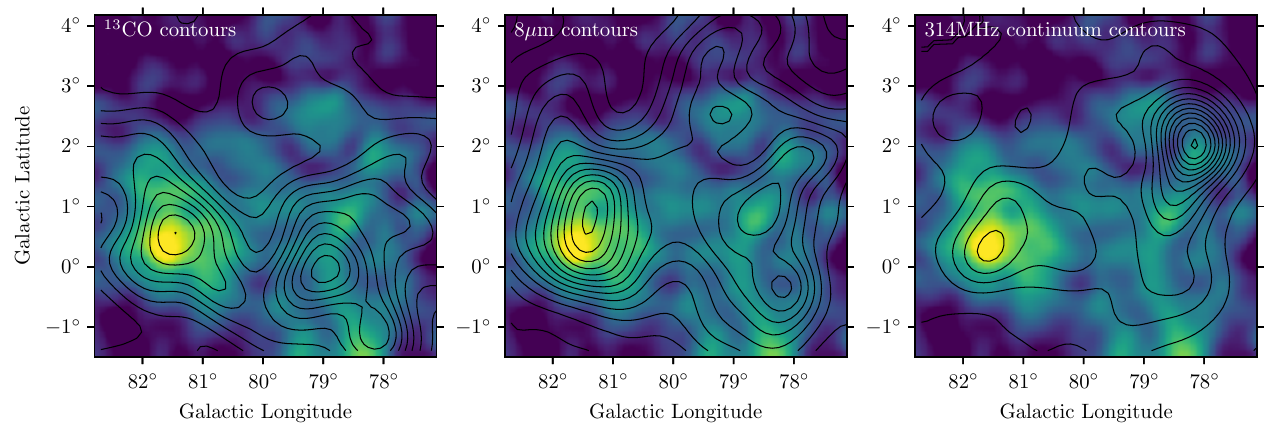}
    \caption{Overlays of the \treceCO{}, 8~$\mu$m, and 314 MHz continuum contours over the \refn{} emission at matched spatial resolution. \treceCO{} contours are drawn at [1, 2, 3,...12]~K~\kms{}. 8~$\mu$m contours are drawn at $[0.50, 0.75, 1.0,...4.25] \times 10^{-3}$~erg~s$^{-1}$~cm$^{-1}$~sr$^{-1}$). 314 MHz continuum contours are drawn at $[60, 120, 180,...1020]$~K.}
    \label{ap:fig:overlay_matched_res}
\end{figure}

We show the comparison of the 8 micron intensity with the \treceCO{} intensity in Figure~\ref{ap:fig:more_fluxcomp}. No strong relation is found.

We show the Point-by-point comparison of the \refn{} with the continuum emission in Figure~\ref{ap:fig:more_fluxcomp}, finding no trends.

\begin{figure}[H]
    \centering
    \includegraphics[width=0.48\linewidth]{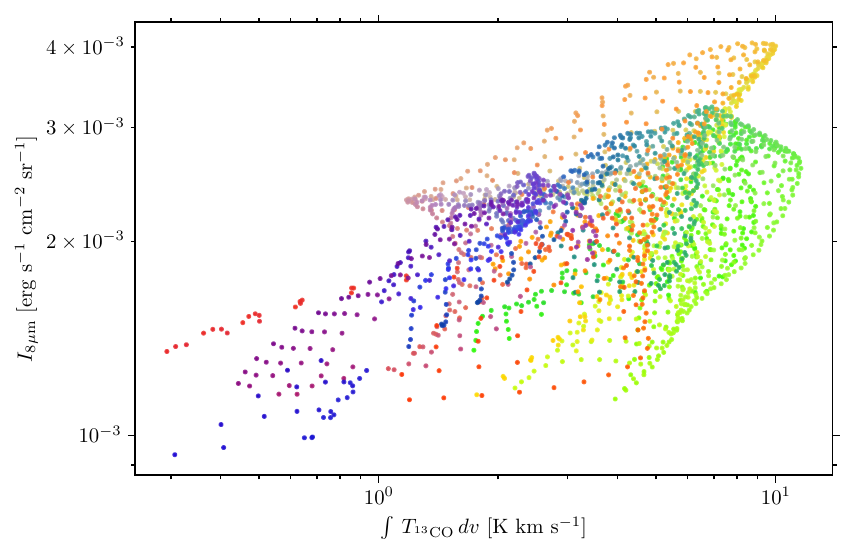}
    \includegraphics[width=0.48\linewidth]{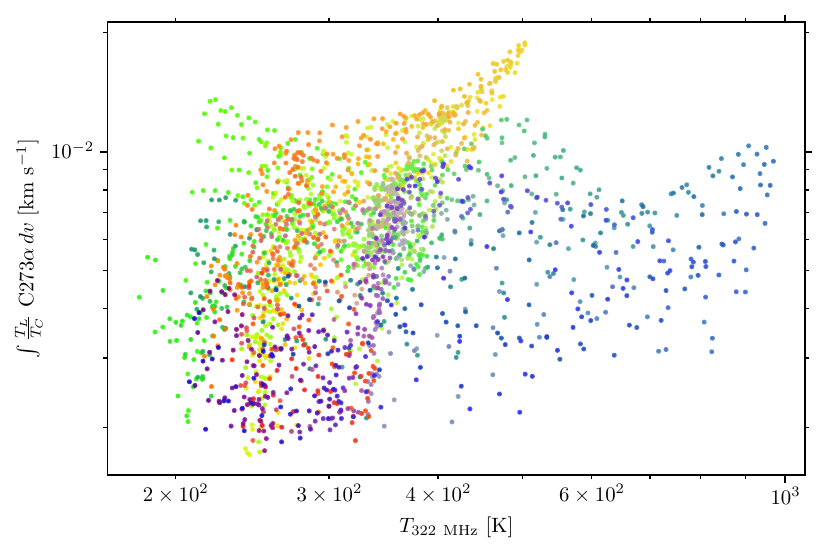}
    \caption{Point-by-point comparison of the 8~$\ mu$m intensity and velocity-integrated \treceCO{} emission \textit{(left)} and a comparison of the \refn{} with the continuum emission \textit{(right)}. The color of the data points are the same as in Figure~\ref{fig:fluxcomp_colorspatial}.}
    \label{ap:fig:more_fluxcomp}
\end{figure}

%%%%%%%%%%%%%%%%%%%
\section{Line fits}
\label{ap:sec:linefitting}

%%%%%%%%%%%%%%%%%%%%%%%%%%%%
\subsection{\refn{} Spectra}

\begin{figure}[H]
    \centering
    \includegraphics[width=0.98\textwidth]{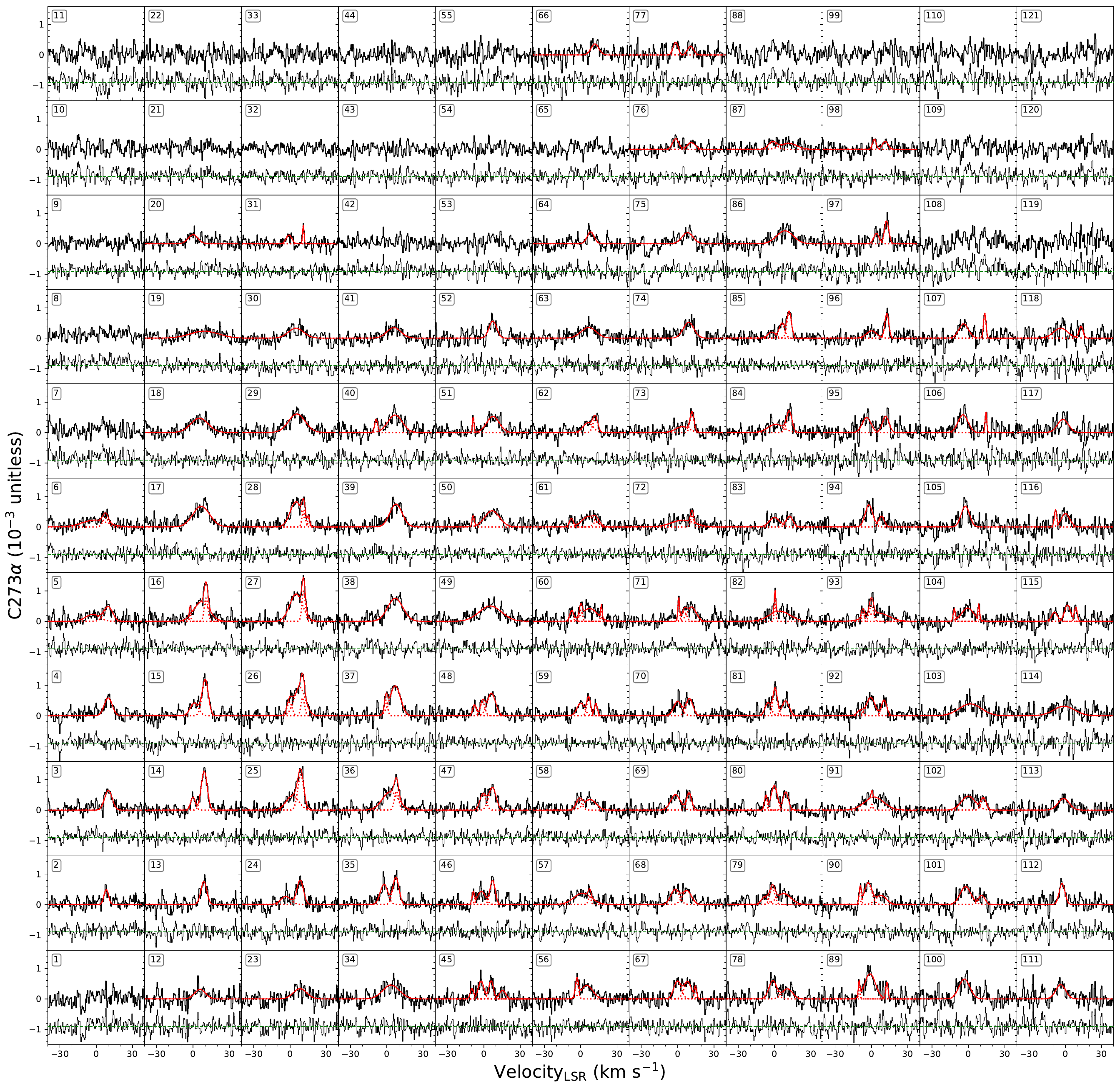}
    \caption{Fits to \refn{} emission. The solid red line indicates (the sum of) the fitted components and the dotted red line indicates the residuals, the data subtracted by the fit.}
    \label{ap:fig:crrl_fits}
\end{figure}

We show the Gaussian profiles fit to the \refn{} spectrum in each aperture region in Figure~\ref{ap:fig:crrl_fits}. The residual spectrum with the fit subtracted from the data is also plotted.

%%%%%%%%%%%%%%%%%%%%%%%%%%%%%%%
\subsection{\treceCO{} Spectra}

\treceCO{} spectra are presented in Figure~\ref{fig:specgrid_multiphase}. We used \texttt{GaussPy+} to fit and decompose the \treceCO{} line emission. Examples of the Gaussian profiles fit to the \CO{} spectrum and the residuals of the spectrum with fits subtracted are shown in Figure~\ref{ap:fig:CO_fits}.

\begin{figure}[H]
    \centering
    \includegraphics[width=0.98\textwidth]{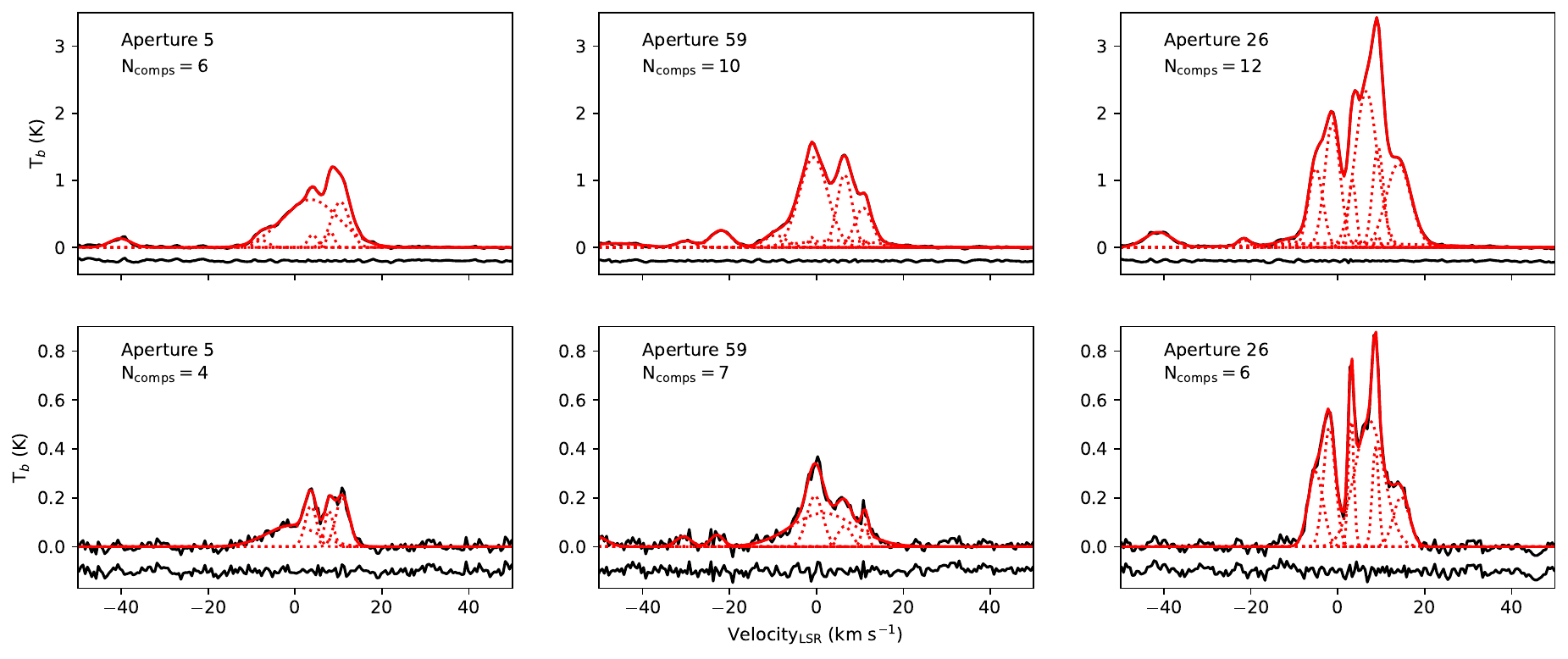}
    \caption{\textit{Top row} is \CO{} spectra and the \textit{bottom row} is the \treceCO{} spectra for the corresponding aperture regions. The aperture ID and the number of Gaussian components fit to the spectra are listed in the top right corner of each plot.
    \label{ap:fig:CO_fits} }
\end{figure}

%%%%%%%%%%%%%%%%%%%%%%%%%%%%%%%%%%%%%%%%

\bibliography{Emig2024_cygx_crrl}{}
\bibliographystyle{aasjournal}

\end{document}